\documentclass[MSNbibl,nameyear,dvips]{arxstspdf}
\usepackage{flushend}
\usepackage{stfloats}
\usepackage{graphicx}

\makeatletter
\def\bsuffix #1{#1}

\setattribute{abstract}{width}{350pt}
\setattribute{keyword}{width}{350pt}

\makeatother

\volume{26}
\issue{4}
\pubyear{2011}
\firstpage{647}
\lastpage{672}
\doi{10.1214/11-STS362}

\begin{document}
\begin{frontmatter}

\title{A Conversation with Eugenio Regazzini}
\runtitle{A Conversation with Eugenio Regazzini}

\begin{aug}
\author[a]{\fnms{Antonio} \snm{Lijoi}\corref{}\thanksref{t1}\ead[label=e1]{lijoi@unipv.it}} \and
\author[b]{\fnms{Igor} \snm{Pr\"{u}nster}\thanksref{t1}\ead[label=e2]{igor@econ.unito.it}}

\runauthor{A. Lijoi and I. Pr\"{u}nster}

\affiliation{University of Pavia and University of Torino}
\address[a]{Antonio Lijoi is Associate Professor of Statistics at the
Department of Economics and Quantitative Methods, University of Pavia,
via San Felice 5, 27100 Pavia, Italy \printead{e1}.}

\address[b]{Igor Pr\"{u}nster is Associate Professor of Statistics
at the Department of Applied Mathematics and Statistics ``Diego De
Castro,'' University of Torino, c.so Unione Sovietica 218/bis, 10134
Torino, Italy \printead{e2}.}

\thankstext{t1}{Also affiliated to Collegio Carlo Alberto, Moncalieri, Italy.}

\end{aug}

% ABSTRACT
%
\begin{abstract}
Eugenio Regazzini was born on August 12, 1946 in Cremona (Italy), and
took his degree in 1969 at the University ``L.~Bocconi'' of Milano. He
has held positions at the universities of Torino, Bologna and Milano,
and at the University ``L.~Bocconi'' as assistant professor and
lecturer from 1974 to 1980, and then professor since 1980. He is
currently professor in probability and mathematical statistics at the
University of Pavia. In the periods 1989--2001 and 2006--2009 he was head of the Institute for
Applications of Mathematics and Computer Science of the Italian National Research
Council (C.N.R.) in Milano and head of the Department of Mathematics at the
University of Pavia, respectively. For twelve
years between 1989 and 2006, he served as a member of the Scientific
Board of the Italian Mathematical Union (U.M.I.). In 2007, he was
elected Fellow of the IMS and, in 2001, Fellow of the ``Istituto
Lombardo---Accademia di Scienze e Lettere.'' His research activity in
probability and statistics has covered a wide spectrum of topics,
including finitely additive probabilities, foundations of the Bayesian
paradigm, exchangeability and partial exchangeability, distribution of
functionals of random probability measures, stochastic integration,
history of probability and statistics. Overall, he has been one of the
most authoritative developers of de Finetti's legacy. In the last five
years, he has extended his scientific interests to probabilistic
methods in mathematical physics; in particular, he has studied the
asymptotic behavior of the solutions of equations, which are of
interest for the kinetic theory of gases. The present interview was
taken in occasion of his 65th birthday.

\end{abstract}

% KEYWORDS
%

\end{frontmatter}

\section{Probability and Statistics at Bocconi University}

\textbf{Antonio:} You received your degree in economics from
``L.~Bocconi'' University in Milano. Why did you decide to study
economics?

\textbf{Eugenio:} I enrolled in an economics degree essentially because
it was the only choice I had. Having attended a Technical High School
for Accountants, at the time the law did not allow students from this
kind of secondary school to study mathematics at university, which
would have been my favorite option. You needed to attend college
preparatory schools to enroll in subjects like mathematics. My family
could not afford my university studies and I~was expected to get a job
right after completing high school. The choice I made at the age of 14
was coherent with this scenario. By the time I got my diploma from high
school, the situation had improved a little bit and I could afford
going to university. This was also due to a financial aid program,
adopted by the farseeing Italian center---left governments of the time,
for students belonging to economically disadvantaged families. The
money from this program (it was around 200 euros, per year, in 1965)
and the earnings deriving from private lessons I used to teach to other
students allowed me to obtain a degree at Bocconi University.

\textbf{Igor:} Why did you choose Bocconi University and not another
university?

\textbf{E:} In the 1960s Bocconi was considered a prestigious
university: a degree from it represented the key for obtaining a good
and rewarding job on the market. These elements exerted an influence on
me and my family. However, there was also a practical reason: Milano is
just one hour by train from my hometown, Cremona, and I could go back
home every weekend. Yet another reason is the influence of my friend
Lorenzo Peccati, a mathematical economist from Cremona, who was a
student at Bocconi while I was still at high school. He was well aware
of my bent for mathematics and suggested a few advanced textbooks where
I started reading the mathematical tools used in economic modeling. In
particular, I~was excited at reading the Italian translation of the
monograph Allen (\citeyear{allen56}) on mathematical analysis for
economists and this convinced me that Bocconi would still have allowed
me to study Maths.

\textbf{I:} This is a funny coincidence since I was convinced by
Lorenzo Peccati's son, Giovanni, by now a well-known probabilist, to
enroll at Bocconi and for precisely the same reasons. Where did your
passion for mathematics come from?

\textbf{E:} I was very lucky at high school because I had a~brilliant
maths teacher, Sidomo Vailati. He had a~variety of scientific and
cultural interests and also did consulting for a few private companies,
thanks to his unusual, at least in that period, knowledge of
statistics, probability and operations research. He certainly was a
self-taught man in the area of Stochastics. At that time, probability
and statistics, unlike analysis, geometry and algebra, were not
perceived as relevant topics within mathematics degrees: they were only
present in a few optional courses. To my knowledge, the only exception
was the University of Roma due to the presence of Bruno de Finetti. In
fact, this unfortunate situation lasted until the 1970s when the first
full professors in probability, apart from de~Finetti, were recruited
after national competitions. Turning back to Vailati, it is worth
recalling that, among some courses for high school teachers organized
by the Italian Ministry of Education, he also took a course in
probability, which was delivered by de~Finetti. As a consequence, at
the age of 16 I was introduced to the realm of subjectivism and learned
the first elements of probability and its applications. These first
years of exposure to de~Finetti's approach have stimulated an
intellectual and scientific interest that has certainly influenced my
later research.

\textbf{A:} How was the environment at Bocconi University in the years
you have been there?

\textbf{E:} Bocconi had very few professors among its own faculty and
heavily relied upon adjunct faculty holding positions in other
universities. These few professors were all influential personalities
of the time, playing significant roles in the Italian social, political
and economic life of the 1960s. For instance, Giovanni Demaria was a
Paretian economist who acted as economic consultant for the constituent
assembly that created the Constitution that lies at the foundation of
the modern Italian Republic after World War II. There was also a
special feeling between Bocconi and Milano, a city that had been able
to overcome the disasters of World War II and was experiencing dramatic
economic growth led by the manufacturing sector. Bocconi looked to me,
and many others, as a vital part of Milano and contributed to
consolidate this process. Then, during the last couple of years, the
student protests of 1968 started. Despite being a private university,
Bocconi experienced serious clashes and some of its students played an
active role in the movement.

\textbf{I:} Did you like studying economics?

\textbf{E:} I was very fond of economics. The professors I~was
interacting with were quite enthusiastic about my inclination toward
developing mathematical tools useful for economic modeling. There is an
episode that occurred during my third year that I like to recall. I~was
attending a course in Political economy which included a series of
seminars and one of them concerned the relationship between the Italian
Central Bank and the Department of Treasury, which at that time was the
subject of a lively debate. For an economic interpretation of the
relationship between the two institutions, we were suggested to refer
to an article by Giorgio La~Malfa and Franco Modigliani; the latter
was later awarded the Nobel prize in economics in 1985. The main
contribution of the paper was the proposal of a static model. Playing
a~bit around with that model, I was able to derive a dynamic version of
it, which seemed in line with the real situation in Italy. This was
appreciated by the other students and the teaching assistants. Also in
connection with the course on Public Finance, I~devised a model
describing the evolution of certain taxing decisions. Overall, I think
I had quite a~good economic intuition.

\textbf{A:} What did lead you to study probability and statistics?

\textbf{E:} I was both impressed and fascinated by the first year
course in mathematics that was taught by Giovanni Ricci. It was more
advanced than a traditional calculus course. The second year maths
course, delivered by Giuseppe Avondo-Bodino, included also a part
devoted to probability, which actually covered essentially the same
material nowadays taught in first probability courses in maths degrees.
During the second year of my degree I also attended a course on
statistical Inference held by Francesco Brambilla, which was important
for my education. Finally, my third maths course by Eugenio Levi
contained some probabilistic applications. This experience revived my
curiosity, dating back to high school, for foundational aspects of
probability. Moreover, I perceived probability as a tough subject and
therefore more challenging and stimulating than others I was studying.

\textbf{I:} What was the topic of your degree thesis?

\textbf{E:} I asked Avondo-Bodino to be my thesis supervisor. He was a
passionate Fisherian and hostile toward the Bayesian paradigm: it
might, thus, seem curious that the title of my thesis was ``The
Bayesian approach to hypothesis testing.'' Indeed, he chose that topic
with the aim of proving the fallacy of the Bayesian paradigm: this is
revealed by the fact that the potential of the Bayesian approach was
going to be assessed with respect to hypothesis testing problems that
had already received well-established answers within the frequentist
framework. To be honest, he did not even like the Neyman--Pearson
approach: according to him, it introduced subjective elements since it
relied on decision theory. My task was essentially to: (i)~collect as
much material as possible on hypothesis testing, (ii)~evaluate the
possible impact of the Bayesian approach and (iii)~establish whether it
could be a sensibile alternative to the frequentist approach. And my
supervisor obviously expected a negative answer to the last question.

\textbf{A:} And how did it work out? What were your first impressions
on the Bayesian approach?

\textbf{E:} While working on the thesis, I developed some skepticism
about the automatic implementation of Bayes' theorem, which was a
legacy from Laplace and his followers. However, my viewpoint was
limited. In fact, writing the thesis was not an easy job, especially
because I could not rely on many systematic and exhaustive treatments.
There were, of course, de Finetti's papers, but, given the unorthodox
way they were written, I was not able to understand the connection
between his theory and the Anglo-American neo-Bayesian approach
typically\break adopted in papers appearing in statistics journals at that
time. De Finetti's work did not follow the standard Bayes--Laplace
paradigm: in contrast, he re-constructed it and recast it in a way to
be coherent with his approach to prediction. The books I~referred to
were Lindley (\citeyear{lindley}), Raiffa and Schlaifer
(\citeyear{raiffa}) and, mostly, Ferguson (\citeyear{ferguson67}),
which contained a beautiful part on the Bayesian approach from the
viewpoint of Wald's decision theory. The 1959 lecture notes of de
Finetti's course at a Summer School in Varenna [later translated in
de~Finetti (\citeyear{defin72})] and Savage (\citeyear{savage}) were
also helpful. I obtained a few minor results in terms of interpretation
and comparison and also derived a ``rule'' for the choice of type-I
error probability $\alpha$. As soon as I completed the thesis, the
monograph DeGroot (\citeyear{degroot}) appeared: I~found it very
interesting and it proved to be very useful for my statistical
education.

\textbf{A:} One of your best friends and main coauthors is certainly
Donato Michele Cifarelli. Did you meet him while studying at Bocconi?

\textbf{E:} Yes, he is actually 10 years older than me and was my
teaching assistant while I was attending the statistics course. He
delivered insightful lectures where it was apparent that he had
remarkable mathematical skills and also a deep knowledge of the book by
E.~Lehmann. Therefore, it was quite natural to seek his help when I
started working on the thesis, which required me to study also the
frequentist approaches. His advice was very important, although at the
time he looked, at least to me, not interested into the frequentist
versus Bayes debate. I was impressed by his vast knowledge of
frequentist methods, both parametric and nonparametric, as well as of
probability theory and stochastic processes. I very much liked the fact
that he preferred fundamental, though maybe difficult, at least for us,
books to much more immediate cookbooks. For instance, he knew in great\vadjust{\goodbreak}
detail the celebrated monograph by J.~L.~Doob on stochastic processes.
We have been bound by a deep friendship and reciprocal esteem ever
since.

\textbf{I:} Who were other important scholars you met at the time and
who influenced your early approach to mathematics?

\textbf{E:} Overall Bocconi was an intriguing place, at least in Italy,
for probability and statistics, given these subjects were, as I said,
almost absent from most maths degrees. At the time Italian statistics,
and academia in general, did not have systematic contacts with the
international community and the head of the Institute of Statistics,
Brambilla, had the merit of introducing and spreading the great
developments in statistics and operations research, which took place in
the UK and the US. He had contacts with foreign scholars: for example,
he is also cited by Leonard J.~Savage in his 1954 book. At the heart of
these scientific activities was the Centre for Operations Research,
which, among its sponsors, could count on Adriano Olivetti, an
enlightened and revolutionary entrepreneur of the time, who ran the
company producing the celebrated typewriter ``Lettera 22'' now
displayed at the Museum of Modern Arts in New York City. His company is also well-known for its pioneering
contributions to the development of personal computers. Among the
various cultural and scientific activities created and supported by
Adriano Olivetti, it is worth mentioning the journal ``Tecnica e
Organizzazione'': Brambilla was a co-editor of the journal when
de~Finetti published an important paper on the essentials of
computational techniques based on Monte Carlo methods, ``Macchine che
pensano e che fanno pensare'' (``Machines that think and that make you
think''). Brambilla was a remarkable figure: he had been the assistant
of Ferruccio Parri who, besides being one of the first Italian Prime
Ministers after the war, was also imprisoned by the Germans during
World War II since he had been one of the antifascist opposition
leaders.

\textbf{A:} What happened after you graduated?

\textbf{E:} I really enjoyed working on my thesis and I was eager to
continue, at least for some time, with research. Ph.D. programs did not
exist in Italy since they were only introduced in the mid-1980s. So
I~was doomed to the military service which was compulsory and would have
lasted for 15 months. I tried to postpone my entry for as long as
possible, since I~wanted to compete for a scholarship from the Italian
Ministry of University. Had I obtained it, I could have freezed it
until the end of the military service. Thankfully\vadjust{\goodbreak} my strategy was
successful and, when I was discharged in January 1972, I was able to go
back to university. After graduation and before starting the military
service, I shared the office at Bocconi with Cifarelli, and together we
attended various maths courses at the State University in Milano. I
then sat the exams during my military service, but in the end I did not
complete a maths degree, since I was already involved in developing my
own research and I was willing to publish! Nonetheless, those studies
turned out to be very useful for me.

\section{From Torino to Bologna, Milano and~Pavia}

\textbf{I:} Unlike many Italian academics, and more in line with what
happens abroad, you have been working in many different universities.
Was this important for your professional development?

\textbf{E:} Definitely. In addition to working in various universities,
I also experienced very different environments. It has been very
helpful from both a scientific and personal point of view. I met many
statisticians and mathematicians with very different backgrounds. My
first experience outside Bocconi was in Torino: it was a small
Department, most colleagues were of my age and so it was pretty easy to
settle in. Afterward I moved to Bologna in a much larger Department
more in line with the Italian statistics tradition. Then I got back to
Milano: first at the Mathematics Department of the State University
and, then, to Bocconi. Finally Pavia, which is one of the oldest
universities founded in 1361 and in a~very prestigious Mathematics
Department of which I am a proud member.

\textbf{A:} Tell us a bit about your first steps in the Italian
academia in Torino.

\textbf{E:} My supervisor, Avondo-Bodino, was full professor in Torino
and a lecturer at his department resigned and decided to leave the
academia. Since they needed a replacement, in December 1973 I moved
there with the concrete opportunity of obtaining, later, a permanent
position. Due to absurd bureaucratic reasons, I finally obtained an
Assistant Professorship only in 1978.

\textbf{I:} You obtained a full professorship position in a~national
competition at a young age in 1980 and, therefore, moved to Bologna
which hosted one of the few faculties of statistics in Italy. Then back
to Milano.

\textbf{E:} That was a time Italy was investing in universities, unlike
now. Therefore, the career perspectives were quite\vadjust{\goodbreak} good also in the
academia if you worked hard. Incidentally, the head of the selection
committee was de~Finetti, I must say a recurrent figure in my life. In
1980 the only autonomous statistics faculties were Roma and Padova,
whereas Bologna and Palermo were offering statistics degrees but within
economics faculties. Statistics became a faculty in Bologna only toward
the end of the 1980s. In fact, and in contrast to what happens outside
Italy, faculties are pivotal players in Italian academia, mainly
because they manage the recruitment. Now it seems that things will
change but, as we say in Italy, everything changes so that nothing
changes. Anyhow, in Bologna I mostly taught probability courses, but my
ties to Milano were still strong, especially because of my
collaboration with Cifarelli. Therefore, I accepted the offer from the
Mathematics Department of the University of Milano in 1984, where they
did not have a faculty member doing research in probability and
mathematical statistics until my arrival. A key role for my transfer
was played by an analyst, Marco Cugiani, whom I also replaced
as director of a research institute of the C.N.R., nowadays the Milano
branch of the Institute of Applied Mathematics and Information
Technology.

\textbf{A:} In 1989 you moved back to your beloved Bocconi University.
Did you find any substantial changes since the last time you had been
there?

\textbf{E:} At Bocconi things had changed a lot, the most apparent
being that it had turned from an elite~to a~larger and more open
university with something like 10,000 students. Hence, I think that
some changes were necessary. As for myself, I was in a~somehow
privileged position since I was mostly teaching advanced and not
compulsory courses, which were much more challenging than most other
courses. Therefore, starting from a yearly basin of more than 2,000
students of very good quality, by self-selection I had small numbers of
students, who were highly motivated and of the highest quality. I guess
I have to mention you two, will not I? But let me also mention Chiara
Sabatti and Giovanni Peccati, among many others.

\textbf{I:} What convinced you to move to another university in 1998?

\textbf{E:} During the 1990s, while I was there, an even more radical
reorganization was occurring: courses of the type I was teaching were
perceived as too ``aristocratic'' and had too few students so that they
were doomed to be shut down. And the same destiny was foreseen for the
most challenging degrees. In fact, when I moved back in 1989, I did it
with the aim\vadjust{\goodbreak} of setting up a statistics degree: I was very disappointed
when the project was officially turned down in 1997. Therefore, I
decided that my experience at Bocconi was concluded and that I wanted
to move back to a maths department. Nonetheless, I still have a special
affective relation with Bocconi.\looseness=1

\textbf{I:} Then you moved to Pavia. And I followed you, given I had
just graduated from Bocconi and started my Ph.D. in Pavia. It was a new
challenge for you at a mature age, was it not?

\textbf{E:} The only science faculty members and future colleagues of
mine in Pavia I knew in person, prior to moving, were three brilliant
mathematicians: Maurizio Cornalba (we were both members of the
scientific committee of the Italian Mathematical Union), Franco Brezzi
(we got to know each other at the meetings of the Italian National
Research Council) and Enrico Magenes. Enrico Magenes, who passed away
last November, shaped the department in its current form and
significantly contributed to its international standing. When I moved,
I was one of only two probabilists and since then we have been able to
hire two additional Assistant Professors. Once arrived, with great
enthusiasm I immediately got involved in the Ph.D. program and the
outcome has been rewarding, as also witnessed by the achievements of
some former Ph.D. students.

\textbf{A:} Throughout your career you have been working at faculties
of science and economics, at public and private universities. What
environment do you think better fits the needs of a researcher in
probability and statistical science?

\textbf{E:} I think that the ideal environment, relatively to the
Italian experience which is the only one I~am aware of, is a maths
department which is open toward the modern trends and research
directions of mathematics and therefore not too much bound to
traditional subjects like analysis, geometry and mathematical physics.
And thankfully there are various departments that comply with these
criteria. However, I must admit I am a bit pessimistic about the
future: the Italian university system in general, and departments
involved in basic research in particular, are struggling and suffering
due to the indiscriminate financial cuts in recent years. These have
been implemented somehow light-heartedly by the government since cuts
in basic research funding are unlikely, at least in Italy, to cause
immediate social upheaval. In fact, it would be important to abandon
the habit of uniformly distributed cuts and aim at creating, or
consolidating, niches of excellence. And I have seen many such niches
during my career.

\section{Bayesian Inference}

\textbf{I:} At the beginning of your academic career you started
working on inferential problems according to the frequentist approach.

\textbf{E:} My interest in frequentist inference started soon after
completing my degree thesis and heavily benefited from the
collaboration with Cifarelli. And one of the first topics we started
working on was hypothesis testing. Corrado Gini and other Italian
statisticians had introduced a considerable number of summary
statistics that were originally used only for exploratory data analysis
to measure, for instance, concentration, variability, dependence and
similarity between sets of data, and so on. Our idea was to use such
summary measures for inferential purposes and, specifically, as test
statistics for studying dependence in nonparametric problems. An early
contribution in this direction was achieved by Cifarelli
(\citeyear{cif75}) who studied the asymptotic distribution of a
statistic arising in a test of homogeneity for two-sample problems.
The paper contained a remarkable result on the distribution of the
integral of the absolute value of the Brownian bridge. Our initial
efforts led to a paper (Cifarelli and Regazzini, \citeyear{cifreg74})
 that we are very proud of: there we determine the
limiting distribution of a measure of monotone dependence introduced by
Gini.
%This paper also paved the way for later developments in \cite{ccr}.
The program we set was very appealing and consisted in checking whether
these statistics, when used for hypothesis testing, yielded tests that
were more efficient than those commonly used at the time. For example,
the index of monotone dependence I was mentioning was compared with
Spearman's $\rho$ and with Kendall's $\tau$ and in some cases it
featured better performances.

\textbf{A:} Around the mid-1970s you turned back to Baye\-sianism. What
about your skepticism?

\textbf{E:} Yes, and my experience at University of Torino was
fundamental in this respect. The department library held the collection
of all de~Finetti's papers, well kept and easily accessible. I started
looking at contributions cited by Savage as decisive for the
foundations of the Bayesian paradigm. My curiosity was fueled by the
fact that, as I said, de~Finetti's work appeared to me completely
disconnected from the Bayesianism I had studied on books and journal
articles. It was a challenging task since de~Finetti's writing style,
which was actually one of the main aspects he was criticized for, was
unorthodox and sometimes seemingly cryptic. Nonetheless, hard work and
stubbornness finally allowed me to understand why de~Finetti
introduced
exchangeability and the role such\vadjust{\goodbreak} a form of symmetry plays in the
reconstruction he gave of the Bayes--Laplace paradigm. This really
opened my eyes on a new world providing a coherent and unified view of
statistical inference, where subjective probabilities play an important
role. In fact, suddenly the subjective interpretation of probability
was the only one that made sense to me from both a philosophical and a
mathematical point of view.

\textbf{I:} You were then able to convince Cifarelli to enter the
realm of Bayesian statistics.

\textbf{E:} I have to say that Cifarelli shared my same doubts on the
foundations of the Bayesian approach to statistical inference. However,
after completing my study program in Torino I pointed him to the
references where de~Finetti was answering our questions and solving our
doubts. Besides de~Finetti's well-known papers, I~had discovered many
other ``minor'' contributions that were important for understanding the
unified framework he had in mind. And, after struggling to understand,
I started to love his style: entering his world had been very
demanding, but once I succeeded the reward was incomparable. In his
work one could find ideas, hints and concepts whose expressive force
was much more powerful than a standard presentation of definitions,
theorems and cool mathematical technicalities. Spurred by the
enthusiasm, I had been able to convince Cifarelli and we started
working together in this direction.

\textbf{I:} Is this when you started your research on Bayes\-ian
nonparametrics?

\textbf{E:} In some sense, yes. On the one hand, we were hoping to be
able to tackle in a Bayesian setting the same issues we had addressed
within classical nonparametric inference. On the other hand, we guessed
that our starting point should have been de~Finetti's representation
theorem as stated in de~Finetti (\citeyear{defin37}) which we could
consider as being nonparametric. In this fundamental paper, the law of
an exchangeable sequence is described as a mixture on a space of
probability measures and the prior is the almost sure limit, in a weak
sense, of the empirical measure generated by the data. This motivated
the investigation of random probability measures (rpm's) for
statistical inference and might have led to extend the Bayes--Laplace
paradigm. We planned to consider estimation of functionals of rpm's
such as the mean, the variance or other characteristic parameters of
the unknown distribution. The necessary preliminary step was to
determine the posterior distribution of these functionals. A helpful
reference was a short paper (de~Finetti, \citeyear{defin35}), where
de~Finetti provides a reformulation\vadjust{\goodbreak} in Bayesian terms of methods used
in exploratory data analysis for smoothing the empirical distribution.
Moving from this, he basically addressed in a nonparametric framework
both the issue of prediction and of evaluation of the posterior
distribution on a set of probability measures. Unfortunately, we had no
clue on how to define a probability distribution on a space of
probability measures that would be analytically tractable. Of course,
we were not aware of T.~S.~Ferguson's paper (Ferguson,
\citeyear{ferguson73}). We were stuck and all the attempts we made led
us nowhere.

\textbf{A:} Was there any decisive event that helped you overcoming
these difficulties?

\textbf{E:} In 1976 I met Andrew~L.~Rukhin who had left the Soviet
Union and was in Italy just before migrating to the US. We discussed
our research activities and I described to him the technical problems
Cifarelli and I were dealing with. He suggested we go through
Ferguson's paper in order to find an answer to our questions. And,
indeed, that was the case: that paper allowed us to resume our project.
So we started considering linear functionals of the Dirichlet process
with the aim of determining their probability distributions
analytically.

\textbf{A and I:} Let us also recall that the study of the Dirichlet
process suited your passion for classical music very well!

\textbf{E:} Gustav Dirichlet is associated with the distribution
because he evaluated the integral on the simplex. The musical
connection is that he married Rebecka Henriette Mendelssohn, younger
sister of Felix Mendelssohn, the famous German composer.

\textbf{I:} Were there other Bayesians in Italy at the time?

\textbf{E:} A few years after its re-flourishing at an international
level, due to the work of Leonard J.~Savage, the Bayesian approach was
sort of rediscovered in Italy as well. This may sound surprising given
de Finetti is Italian: however, one has to consider that de Finetti
only entered academia in 1946, at the age of 40, when his research was
already focused on different topics. Interestingly, he had obtained the
position already in 1939, but could only start his job in 1946 after
the fall of the fascist regime due to a law forbidding the appointment
of unmarried professors, as was de Finetti's case.
Anyhow, in those years there was a large
group led by Giuseppe Pompilj in Roma and some scholars started to work
on Bayesian statistics, like Ludovico Piccinato. In Roma there were
also some of de~Finetti's students like, for instance, Fabio
Spizzichino. I should also mention a group based in Trieste and
coordinated by Luciano Daboni, who started working\vadjust{\goodbreak} under de Finetti's
supervision soon after gaining
his university degree.
Besides actuarial mathematics, they focused mainly on exchangeable
processes and foundational issues of Bayesian inference and, during the
years, I had many fruitful interactions with them.

\textbf{A:} Even if more interested in the Bayesian para\-digm, you did
not avoid doing research based on a~frequentist approach. It seems you
did not, and still do not, see any ideological contraposition between
Bayesianism and frequentism.

\textbf{E:} I have never seen this as an ideological contraposition. I
think that ideological positions make sense only outside the realm of
mathematics. Anyhow, even when I was working on statistical problems
according to the frequentist approach, I always had the feeling that
the Bayesian framework was far more complete and logically sound. I was
not enthusiastic about the automatic use of priors on unobservable
parameters: the subjective views I had on probability were in conflict
with such a treatment of the Bayes--Laplace paradigm, as I believe that
inference must concern quantities that can be empirically observed.
But, on the other hand, the Fisherian attitude appeared to me as too
drastic, because prior beliefs should play a role in statistical
inference. Once able to fully understand the consequences of
de~Finetti's results, I became convinced that Bayesianism was the only
acceptable way of inductive reasoning.

\textbf{A:} Current developments in Bayesian inference involve a heavy
use of simulation algorithms. Do you still think there is a need for
putting a strong effort in determining exact forms of Bayesian
inferences (or, at least, error evaluation when approximations are
used), even when these are difficult to use in practice?

\textbf{E:} Computational techniques have been decisive in making
Bayesian models applicable to real world problems and some recent
applications I saw are simply amazing. I definitely think that the
advantages they yield largely surpass some drawbacks associated with
their uses. That said, I would still like to make a point, which I
think is important since it has to do with how statistical modeling is
conceived. Indeed, models should be devised as simple as possible,
while still preserving the capability of capturing the essential
features of the phenomenon under study. Such a simplification could be
achieved by first detecting inessential elements and, then, dropping
them when it comes to the point of specifying the model. This attitude
is natural when one aims at achieving exact estimates of the quantities\vadjust{\goodbreak}
of interest. However, if the need for pushing analytic results as far
as possible disappears, it is likely that the models become more loose
and unnecessarily complex. Both parsimony and extreme care in the
formalization of models are still important guidelines for research:
the only difference is that they now need to be spelled out clearly,
while they were implicitly followed in the past. Another related and
important point concerns approximation. When exact inferences are not
possible, one should put some effort in providing an upper bound to the
error of approximation yielded by the numerical techniques that are
used. I have tried myself to work in this direction, for instance, in
relation to approximating the probability distribution of the mean of a
Dirichlet process. I know this is a challenging task, but it cannot be
avoided.\vspace*{3pt}

\section{de Finetti and the Influence of de~Finetti's Work}\vspace*{3pt}

\textbf{I:} There is no doubt your research has been deeply influenced
by de~Finetti's work. Which was the first paper of de~Finetti you read
through?

\textbf{E:} While I was completing my thesis at Bocconi I~came across
his joint paper with Savage (de~Finetti and Savage,
\citeyear{definsav62}). It contained a discussion on the choice of the
prior distribution and was mainly illustrative with no deep mathematics
involved but still evocative for a novice.

\textbf{A:} His most renowned piece of work certainly is the two-volume
book on probability theory, de~Finet\-ti (\citeyear {defin70}). What else
would you suggest to a student who is willing to study and understand
de~Finetti's stance in probability and statistics?

\textbf{E:} I would certainly suggest de~Finetti (\citeyear {defin06}),
two volumes containing selected papers by de Finetti, which have been
published by the Italian Mathematical Union in 2006 in occasion of the
centenary of his birth. The first volume is on probability and
statistics, whereas the second is on applied maths and on the teaching
of maths. As for his subjective views on probability, one should refer
to de~Finetti (\citeyear{defin31}). One should also read de~Finetti
(\citeyear{defin37}). Another important piece of work is de~Finetti
(\citeyear{defin72}). Finally, de~Finetti (\citeyear{defin92}) contains
a selection of some of de~Finetti's papers with English translation.
Unfortunately, some significant contributions, at least to my
knowledge, have been only published in Italian, such as those related
to independent increments processes and some others on the
subjectivistic definition and interpretation of probability.\vadjust{\goodbreak}

\textbf{I:} As you just mentioned, the fact that he was not writing in
English hindered the circulation of his ideas and results in the
scientific community.

\textbf{E:} This is definitely true. For example, it is probably
unknown to many that de~Finetti introduced the celebrated $\tau$ index
a few years before Kendall (de~Finetti, \citeyear{defin37b}). In 1939
he obtained some important results on optional stopping: indeed,
de~Finetti (\citeyear{defin39}) deals with the gambler's ruin problem,
where one can also find an embryonic version of the Girsanov theorem.
Another important contribution was the continuity theorem for
characteristic functions: he proved it in the appendix of de~Finetti
(\citeyear{defin30}). Besides these, it is worth listing a few other
contributions for which a priority to de~Finetti should be
acknowledged: he completed what is now known as the Glivenko--Cantelli
theorem before Francesco P.~Cantelli in de~Finetti
(\citeyear{defin33}); in de~Finetti (\citeyear{defin40}) he devised a model that anticipated the portfolio theory for
which Markowitz was awarded the Nobel prize; he proved the theorem on almost
everywhere nondifferentiability of the trajectories of the Brownian
motion in de~Finetti (\citeyear {defin29}).

\textbf{A:} With reference to the de~Finetti (\citeyear {defin29})
paper, which is actually our favorite, should we not as Italians
propose \textit{L\'{e}vy processes} be called \textit{de
Finetti--L\'{e}vy processes} instead?

\textbf{E:} As I mentioned before, the answer is affirmative. Indeed,
de~Finetti started from a more general problem of providing the random
counterparts of a Volterra classification for the ordinary laws of
physics. In this context he identified processes with independent and
homogeneous increments as those whose characteristic function satisfies
the first of the equations in Volterra's classification, namely,
$X'=f(\lambda)$. As a by-product, he also introduced implicitly the
notion of infinite divisibility. In a subsequent paper, de~Finetti
(\citeyear{defin30b}), he further characterized the class of infinitely
divisible laws as the class of distribution limits of compound Poisson
processes, thus providing a representation theorem for infinitely
divisible distributions. L\'{e}vy was not aware of de~Finetti
(\citeyear{defin29}) and resorted to a different approach to obtain
more general and deep results. The contribution by Khintchine to the
well-known L\'{e}vy--Khintchine representation originates from a paper
published in 1937 (see Khintchine, \citeyear{khintch37}): Khintchine's
paper builds upon Kolmogorov (\citeyear{kol32}), where Kolmogorov
explicitly mentioned (even in the title of the article) that he was
resorting to the approach set forth by de~Finetti. So, yes, it should
definitely be \textit{de Finetti--L\'{e}vy processes}.\vadjust{\goodbreak}

\textbf{I:} And what were his connections with the broader
international scientific community?

\textbf{E:} His first international contacts, before graduating in
mathematics at the University of Milano, are related to a paper,
de~Finetti (\citeyear{defin26}), he wrote on Mendelian inheritance,
which had quite an impact in biology. It was his first paper and
appeared on the   Italian journal
\textit{Metron}. His results also attracted the attention of Alfred J.
Lotka and Jacques S. Hadamard. The latter was so impressed by
de~Fi\-netti's achievements that he suggested Georges Darmois to study
the paper, as witnessed by one of the letters that de~Finetti wrote to
his mother in 1929 and that have recently been published by his
daughter Fulvia. This research also originated the so-called de~Finetti
diagrams that are extensively used in population genetics.

\textbf{A:} An important event at which de~Finetti drew attention on
his research in probability was the International Congress of
Mathematicians (ICM),\break which was held in Bologna in 1928.

\textbf{E:} That conference was definitely important for the
development of de~Finetti's interactions with foreign scholars. On that
occasion he presented his first results on exchangeability and made
contact with Maurice R. Fr\'{e}chet, who later invited him to the
Institut Henri Poincar\'{e} in 1935 and to the Colloque de
G\'{e}n\`{e}ve in 1937 where he then also met Jerzy Neyman and others.
He had frequent interactions with Paul L\'{e}vy and Aleksandr
Khintchine, respectively, on independent increment processes and on the
proof of the representation theorem for exchangeable sequences. He was
also in contact with Andrey N. Kolmogorov, as witnessed by the
Kolmogorov (\citeyear{kol32}) paper on infinite divisibility whose
title contains an explicit reference to ``A problem of de Finetti.''
Both Kolmogorov and de Finetti also worked at the same time on the
derivation of a~representation theorem for associative means, now known
as the de~Finetti--Kolmogorov--Nagumo Theorem. He also got in contact
with many eminent mathematicians via mail. In fact, he used to have a
notebook in which he recorded to whom he had sent which of his papers:
de Finetti's daughter, Fulvia, once showed it to me and the names are
impressive. After World War II he had significant scientific
collaborations with Leonard J. Savage and Lester Dubins and he
interacted also with William Feller and Abraham Wald.

\textbf{I:} Were his views on the subjective approach to probability
theory held in high regard?

\textbf{E:} In mathematics his work has been largely\break ignored, and not
only because of the subjective interpretation. Indeed, the mathematical
approach\break yielded by such interpretation does not require
$\sigma$-additivity. In fact, finitely additive laws also become
admissible and the traditional measure--theoretic approach to
probability theory represents obviously a~particular case. Countably
additive probabilities are coherent in de~Finetti's sense but are just
a subclass of coherent laws. And de~Finetti himself was well aware that
many results could have been neater by assuming countable additivity.
We may reasonably conjecture that his position in favor of including
finitely additive probabilities somehow put him off from focusing on
the particular countably additive setup. This could explain, for
example, why he did not further investigate processes with independent
increments. It is to be noted that the framework for his subjective
approach had been settled by 1931 and, as evident from his published
mail exchange with M. Fr\'{e}chet, he fought for it for a while.

\textbf{I:} And what about the impact on statistical practice?

\textbf{E:} In Bayesian statistics references to subjectivism are quite
frequent, but I actually see little of de~Finet\-ti behind them. First,
in the subjective approach also finitely additive laws are allowed and,
therefore, a~proper subjectivist should try to analyze statistical
problems in this setup. This point is very important in the case where
``transcendent'' conditions---such as convergence of sequences of
random elements, forms of the corresponding limits, etc.---are
involved: one should, then, establish the extent to which the
conclusions depend on the specific $\sigma$-addi\-tive extension (usually
unique) of the original finite-dimensional distributions. Second, from
an interpretation point of view, subjectivism and objectivism are often
mixed up and Bayes theorem is applied in an automatic way, whereas
subjectivism would require probabilistic statements to be made on
verifiable events. Subjectivism seems more a kind of catch-phrase than
a real commitment. In my opinion, the papers of L.~J.~Savage,
L.~Dubins, J.~Pitman, P.~Diaconis and D.~Freedman are the ones that
adhere most closely to de~Finetti's views.

\textbf{A:} Did your convinced support of subjective probability affect
the way you teach probability courses?

\textbf{E:} This represented a sort of dilemma throughout my career.
Focusing solely on de~Finetti's mathematical theory of probability
would have implied providing students with an unorthodox background in
probability: it could have been an enrichment for some of them but also
a drawback for some others, especially for those who needed to use
probability as a mere tool in other disciplines. Therefore, most of the
courses I taught were within the $\sigma$-additivity framework.
Nonetheless, I have always tried to illustrate extensively some
distinctive features of the subjective viewpoint in one of my first
lectures. This was useful since it provided students with a more
complete picture of the subject and allowed them to understand that the
results I was going to state and prove were valid on a special class of
probabilities sharing the property of countable additivity. Students
were, then, aware that it was somehow like teaching them a course in
analysis that was just about analytic functions! The connection with
conditional properties was far more difficult to point out. As for the
subjective interpretation, it is still possible to preserve it even
when confining to $\sigma$-additive probabilities.

\textbf{A:} Can you provide some further insight on this last issue?

\textbf{E:} The difficulty I am referring to arises due to the fact
that Kolmogorov's definition cannot be seen as a special case of
coherent conditional probabilities. In fact, the Kolmogorov approach lacks an
appropriate axiomatization and interpretation of conditional
probability: the definition is by means of a limiting procedure. The
perspective is then completely different.
For example, de~Finetti's approach necessarily leads to conditional
probabilities that are regular and proper, whereas it is well known
that Kolmogorov's definition does not. In order to grasp these
mathematical and conceptual differences on conditional expectations and
probabilities, one can refer to the works by L. Dubins, David Blackwell, Czeslaw Ryll-Nardzewski, William
D.~Sudderth, Ro\-ger~A.~Purves, Pietro Rigo, Patrizia Berti and also
myself.

\textbf{I:} Tell us about your meetings with de~Finetti.

\textbf{E:} I first met him in 1969 at a summer course~on mathematical
economics in Urbino. Since I was working on my thesis, I took the
opportunity to ask~him a few questions about his paper with Savage
(de~Fi\-netti and Savage, \citeyear{definsav62}) I had read. He was not
very talk\-ative and probably thought I was not understan\-ding anything.
He was right, but I still went away with the impression that it was not
simple at all~to~in\-teract with him. Afterward I met him at some
confe\-rences during the 1970s, but at that time he was not working on
statistics and probability with the same intensity and creativity of
the early days: he was more inclined to elaborate on general
philosophical and foundational aspects. The only thing I can say about
our meetings is that I had the impression he was interested in
nontrivial and original approaches or attitudes that to some other
people might have appeared as singularities. For example, in Bologna he
once told me he had been fascinated by the mathematical physics
lectures held at the Polytechnic in Milano by a lecturer, Bruno Finzi,
whose assignments were notoriously challenging and contained exercises
that Finzi himself could not solve. He recalled the solutions he had
been able to give were very original and much appreciated by Finzi. He
also told me he had appreciated lectures on economics of insurance
companies delivered by Ulisse Gobbi at the Polytechnic in Milano
because they had been the source of inspiration for the mathematical
modeling of many aspects of economics he had later investigated. I am surprised by this, since in Gobbi's work I did
not find any mathematical formalism.\looseness=-1

\textbf{A:} He was also engaged in public life and gained some
popularity because of his political experience.

\textbf{E:} His political experiences can be well understood if one
refers to the environment where he grew up. De~Finetti's family was
wealthy and highly educated. They were part of the Italian community in
territories of the Habsburg Empire, and his father was an engineer
working for the Austro-Hungarian railway. During his childhood he had
learned about the irredentist ideas of the Italian minority that was
aiming at unification with Italy. Such aspirations quite naturally
developed into strong nationalist feelings once the area became part of
Italy. Moreover, having been part of a minority, he developed a strong
sensitivity toward injustice in all respects and, therefore, also a
strong criticism toward some social implications of capitalism of the
time. This blend of ideas somehow naturally led him to support the
rising fascist party: its initial political and social program included
a series of reforms whose goal was the complete State control of the
economy. As de~Finetti himself wrote a few years before dying, the
direction of the whole economy, once freed from the terrible tangle of
individual and interest group selfishness, should lean toward the
collective achievement of a~Paretian ``optimum'' and should be further
inspired by ``fairness'' criteria.

\textbf{A:} Hence, his support to fascism was mainly the result of
ideal feelings that were fueled by strong social and economic views.

\textbf{E:} This is further witnessed by the fact that after the fall
of fascism, he sympathized with left-wing movements without adhering to
a large political party. Finally, during the 1970s he started being
involved in important campaigns for civil rights and for social
justice. The Italian party that better fitted his political thoughts of
the time was the Radical party.

\textbf{I:} Can you tell us something about it? It seems that, while
being involved in political activities set forth by the Radical Party,
he spent one night in~jail!

\textbf{E:} In fact, he did not end up in jail because the order to
release him
arrived before being imprisoned. To make a long story short, he was
editor of
a newspaper of the Radical Party, which was publishing letters of conscientious
objectors who refused to perform the compulsory military service. This
was illegal
at the time. The day he learned he was going to be arrested, he asked
the police
whether it was possible to arrest him at the \textit{Accademia dei
Lincei}, the
most prestigious Italian science academy, where he was going to have an official
meeting the day after. He motivated such a seemingly bizarre request
with the fact
that the police could have saved some money by not picking him up by
car at home:
the \textit{Accademia dei Lincei} building was, indeed, just a~few
steps away from
the prison he was supposed to go to. However, the order to release him
arrived as
soon as he got to jail. This episode had a huge echo in the press.
%See, e.g, \cite{fulvianicotra} for further details on this episode.

\textbf{A:} We also recall a story you told us about Kolmo\-gorov
visiting Roma and wanting to meet de~Finetti.

\textbf{E:} In 1962 Kolmogorov was awarded the Balzan prize for
Mathematics, the other awardees being\break Pope Giovanni XXIII for Peace,
Paul Hindemith for Arts, Samuel E. Morison for Humanities and Karl von
Frisch for Biology. Two well-known mathematicians, Gaetano Fichera and
Olga A.~Oleinik, went to collect him at the Roma airport and asked him
what they could do for him. And, as Fichera reported, his answer was,
``If you know him, then you should organize a meeting with de
Finetti.''

\textbf{A and I:} De Finetti's papers are scattered with brilliant
ideas, sometimes only sketched. What are the aspects of de Finetti's
work which still need to be developed?

\textbf{E:} As for some specific topics, such as exchangeability and
processes with independent increments, in my opinion most of his ideas
have already been extensively developed and not much is left to
investigate in the precise direction he had originally thought of.\vadjust{\goodbreak} On
the other hand, I believe that much is still left to investigate on the
general foundations of probability theory that emerge from his work and
that he strongly supported. These studies might have a relevant impact
in statistics, in physics and in other research areas. The advances I
am thinking of concern both the interpretation of probability and the
enlargement---along with its mathematical impli\-cations---of the class
of admissible probability laws to include also the finitely additive
ones.

\section{Probability and Statistics in Italy}

\textbf{A:} You investigated quite extensively the development of
statistics and probability in Italy in the first half of the 20th
century (e.g., Regazzini, \citeyear{reg05}). Can you tell us about it?

\textbf{E:} In contrast to what happened in the Anglo--American world
or in Russia, in Italy probability and statistics developed along
almost separate paths. Probability started growing in mathematical
environments. As far as I know, the first to deal with the topic in a
comprehensive way was Guido Castelnuovo, a famous mathematician who was
mainly doing research in algebra and geometry. His 1919 book on
probability (Castelnuovo, \citeyear{castel}) was used as a~textbook for
quite some time in those few mathematics degrees where probability was
taught. The interpretation of probability was frequentist, in line with
a view that would have been later shared also by Fr\'{e}chet, L\'{e}vy
and Kolmogorov, and covered results of the Russian school up to Andrey
Markov and Aleksandr M.~Lyapunov. Already, back in 1915 he had the idea
of setting up a school of statistics and actuarial sciences at the
University of Roma, which was then created in 1927. It had considerable
success with many enrolled foreign students and then became a proper
faculty in 1936 with Gini. In the preparation of his book Castelnuovo
was helped by Cantelli, who is considered, also at an international
level, one of the first modern probabilists. He derived, among other
contributions, versions of the laws of large numbers, the
Borel--Cantelli lemma, a~mathematical theory of risk that was named
after him, and developed an autonomous abstract meas\-ure--theoretic
theory of probability, which appeared before Kolmogorov's. It is
interesting to recall that in this last development a crucial point was
the proof of the existence of measurable maps defined on $[0,1]$,
endowed with the uniform distribution, in such a way they have
prescribed probability laws: such an approach also reflects the idea of
adhering to the classical definition of probability due to
Laplace.\vadjust{\goodbreak}
Anyhow, this problem led him and his students to anticipate at least
part of what is nowadays known as the Skorokhod
representation. A~distinguished scholar who obtained important results
along the lines of research undertaken by Cantelli was Giuseppe
Ottaviani, who is also known for his inequalities that are related to
Cantelli's theory of risk. Francesco G.~Tricomi, eminent analyst
and\break
friend of Cantelli, also gave some contributions to probability as did
Carlo E.~Bonferroni, who is well known for his inequalities.

\textbf{A:} Given such a glorious tradition, it is quite surprising, as
you said earlier, that the first full professors in probability were
appointed by Italian universities only in the 1970s, with the notable
exception of de~Finetti.

\textbf{E:} Actually, at the beginning of the 1970s only two professors
in probability were recruited, namely, Giorgio Dall'Aglio and Giorgio
Letta. Dall'Aglio was at the Faculty of Statistics in Rome and was a
member of the before mentioned group led by Pompilj. Letta is from Pisa
and spent several research periods in Germany and France. The latter
experience stimulated collaborations between Italian probabilists---some of whom were Letta's students---and French probabilists in Paris
and Strasbourg, a fruitful trend which is still ongoing. Then a larger
group of people were appointed at the end of the 1970s in various
Italian universities.

\textbf{I:} And what about statistics?

\textbf{E:} In the last three decades of the 19th century, topics that
are today ascribed to Mathematical Sta\-tistics were taught in geodesy or
astronomy courses. Lectures by a not well-known Italian mathematician,
Paolo Pizzetti, were very interesting and contained some innovative
ideas on significance tests. More conventional, at least according to
the Italian framework, statistics courses were in law faculties: many
academic statisticians actually had a~degree in law. Most of them were
involved in Official statistics and it was therefore natural that the
interactions between statisticians and probabilists were rather
limited. The first modern Italian statistician was Rodolfo Benini, who
had a law degree from the University of Pavia and developed statistical methods for demographic,
sociological and economic problems around the end of the 19th and the beginning of the 20th
century. I recall once I came across historical documents presented in
noteworthy conferences of the American and British Economic Societies
where Benini is referred to as one of the founders of econometrics. I
think this due to his analysis of\vadjust{\goodbreak} income and wealth distributions and
to the pioneering use of multiple regression methods to estimate, for
example, demand curves. He also had the idea of studying contingency
tables with fixed marginals. Among his successors, the main figure is
certainly Gini, also a graduate in law. His methodological
contributions to statistics were praiseworthy and were later studied
not only in relation to mere data analysis. Gini dominated Italian
statistics until his death in 1965 and created a school of faithful
followers. A prominent group of scholars was led by Pompilj at the
Faculty of Statistics in Roma. As I~recalled earlier, Dall'Aglio was
one of its members and he obtained noteworthy mathematical results that
can be traced back to the Ginian analysis of statistical relationships.
His results, however, have a remarkable independent interest: for
example, he provided a relevant contribution to the definition and to
the properties of what is today known as the Wasserstein distance. See
Dall'Aglio (\citeyear{garlic56}).
%Finally, it is also important to mention Paolo Pizzetti, who was
%active at the end of the 19th century and proposed independently what
%is now known as Pearson's $\chi^2$ goodness of fit test.

\textbf{A:} You mentioned Paolo Pizzetti who seems to be a neglected
figure within the Italian statistics community, was he not? We have
never heard of him in our statistics courses.

\textbf{E:} Yes, he was unfairly neglected. His contributions, which appeared in the 1880's, were very
innovative and relied on an original approach that somehow anticipated
a few distinguishing ideas lying at the foundations of statistics as
set forth by Karl Pearson and by Ronald A.~Fisher. As an example, he
proposed procedures that were very similar to the significance tests
Fisher would have later adopted as a distinctive feature of his
methods. Pizzetti also had remarkable mathematical skills that allowed
him to determine the exact distribution of certain statistics used for
data analysis. And he was well aware of the results achieved, in this
direction, by a German geodesist Friedrich R. Helmert. He reproved
Helmert's results with the aim of extending them and relied on
innovative methods and techniques that Fisher himself would have later
proposed independently. This is very well documented in a recent
historical monograph by Anders Hald. As you can easily guess,
Pizzetti's ideas were totally different from those that Gini would have
later expressed apropos of the Fisherian tests. Indeed, Gini was very
critical about the use of significance tests and his criticisms were
shared by de~Finetti. This may partly explain why Pizzetti is not known
by many statisticians. His work was somehow considered as heterodox for
quite some time, as demonstrated by the 1960s reprint of Pizzetti's
1892 book (Pizzetti, \citeyear{pizzetti63}).\vadjust{\goodbreak} In the preface, written by
V.~Castellano, P.~Fortunati and G.~Pompilj, it is claimed that parts of
Pizzetti's work were ``misleading and\ldots con\-tained errors that had
been masterfully pointed out by Gini in Gini (\citeyear {gini39}).''
And the ``misleading parts'' they were referring to are exactly those
where Pizzet\-ti uses his results for devising statistical tests.

\textbf{A and I:} The excerpt you read can partly explain the isolation
of the Italian statistics community in those years.

\textbf{E:} It partially does. Indeed, I think that Gini's critical
remarks make sense. The point is that they were not complemented by
alternative proposals that could take his concerns into account.
Hence, it was almost inevitable that Gini's position would have become
marginal and isolated within the broader international community. It
should be recalled that isolation fitted very well with the political
climate favoring autarkic tendencies during the fascist regime and it
unfortunately further consolidated over the years in the Italian
statistics community, at least in academia. This obviously had a
long-lasting negative impact, from which Italian statistics started recovering
only in the 1970s.

\textbf{A:} Gini was appreciated both for his scientific achievements
and for his praiseworthy services as a scientific expert within various
important Italian institutions.

\textbf{E:} Definitely. He was founding President of the Italian
Central Institute of Statistics in 1926 and set up first the School and
then the faculty of statistics, demographic and actuarial sciences in
Roma in 1936. He was in constant contact, also meeting him in person,
with Mussolini, who used to pay attention to statistical analyses for
taking decisions on policy issues. For example, he acted as a technical
advisor within the programs of demographic and eugenics policies
pursued by the fascist regime. Later he also  founded the Italian
Statistical Society, of which he has been President for 20 years.
Besides the scientific and institutional authoritativeness he gained in
Italy, it should be recalled that he obtained countless recognitions
abroad as well. Among them I could mention that he became Honorary
Fellow of the Royal Statistical Society, Vice President of the
International Sociological Institute, and Honorary Member of the
International Statistical Institute. In 1920 he was the founding Editor
of the journal \textit{Metron}, which published papers by many eminent
statisticians of the time, such as R.~A.~Fisher, A.~A.~Chuprov,
A.~J.~Lotka, S.~S.~Wilks, E.~E.~Slutsky, S.~Kullback, H.~Wold and
A.~L.~Bowley.

\textbf{I:} We have also heard of some funny stories about Gini bearing
ill-luck. Can you tell us something more?

\textbf{E:} Yes, this is somehow true, but it is to be considered
within the typical Italian attitude of making fun of powerful people,
as Gini certainly was. There are various minor anecdotes and a dramatic
episode that would allow to conjecture a ``correlation'' of the type
you are referring to. As for the latter, something incredible happened
in 1927: he was on the steamboat ``Princess Mafalda,'' which
shipwrecked off the Brazilian coast between Salvador de Bahia and Rio
de Janeiro, and he was among the few survivors, the ``legend'' says
thanks to his rowing skills, a sport he had practiced in youth. A less
dramatic and funnier story I have heard of concerns an episode where,
chatting with a colleague of his, he paid a compliment to a young
female student's legs whom they met on the stairs: after a few steps
she fell down and broke her leg. I remember that Ottaviani did not
mention his name, he referred to him as \textit{the unnamed}, since
mentioning his name could have led to something bad happening. All
kidding aside, after the shipwreck in Brazil, he criticized the Italian
authorities for the poor assistance from the Italian Navy and, more in
general, from the Italian government. These complaints caused him a lot
of troubles with the fascist regime in Italy. He had, indeed, a~strong
and straight attitude that helped him to protect scientific matters and
appointments from political influence. Of course, this position
attracted the aversion of many Fascist party officials who strove for
Mussolini to remove him as president of the Italian Central Institute
of Statistics. And his criticisms on the occasion of the shipwreck were
added to the list of Gini's ``offences'' to the regime that led to his
resignation in 1932. However, as I~said before, he kept collaborating
with the regime as a~scientific expert in demography, statistics and
eugenics.
%A rich and detailed account of these episodes, along with a
%description of Gini's relationship with the fascist regime, can be
%found in \cite{cassata}.

\textbf{A:} Cantelli, de~Finetti and Gini were the towering figures in
probability and statistics before World War II in Italy. They were also
completely different characters. How did they get along?

\textbf{E:} Gini published de Finetti's work on Mendelian inheritance
in \textit{Metron} and offered de Finetti a job at the Italian Central
Office of Statistics before he graduated. While at the Italian Central
Office, de Finetti was involved in a project for predicting the
evolution of the Italian population and crucially designed all modeling
aspects of the project. He then wanted this to be credited as his
contribution, but Gini was reluctant to do so. This episode is well
documented in one of the letters de~Finetti wrote to his mother and
contained in the collection published by his daughter that I have
already mentioned. In any case, at the end of his four year contract in
1931, de Finetti moved back to  Trieste and started to work
for the insurance company \textit{Generali}. The relationship between
de Finetti and Cantelli was quite a~difficult one, since they were in
strong disagreement on the interpretation of probability. Cantelli did
not want to hear anything about finite additivity and he also tried to
prove that $\sigma$-additivity was a necessary property.

\textbf{I:} In addition to \textit{Metron}, there was also the
\textit{Giornale dell'Istituto Italiano degli Attuari} (\textit{GIIA}), which
was a top journal in statistics and probability during the 1930s. Why
have they both lost their international reputation since then?

\textbf{E:} The \textit{GIIA} was established in 1930, the same year
\textit{The Annals of Mathematical Statistics} published their first
issue. It was edited by Cantelli and the most distinguished scholars of
the time, such as Cram\'{e}r, Fr\'{e}chet, Kolmogorov, Khintchine,
L\'{e}vy, Neyman and von Mises, published fundamental contributions on
it. World War II ruined everything, since its publication was suspended
and the \textit{GIIA} lost its elite status among the top probability
and statistics journals which, during and soon after the war, included
\textit{The Annals of Mathematical Statistics}, along with
\textit{Biometrika} and the \textit{Journal of the Royal Statistical
Society}. The other Italian prestigious journal, \textit{Metron}, which
was established in 1920, paid a high price for the line of development
of Italian methodological statistics and actually already declined
before the war.

\textbf{A:} In 1978 you actually published a very interesting paper
characterizing the Dirichlet process in terms of linear predictive
distributions (Regazzini, \citeyear{reg78}) on the \textit{GIIA}. Why
did you decide this was a~suitable outlet for your paper?

\textbf{E:} After my discussions with Rukhin, the Dirichlet process
became a main ingredient of my research agenda. In fact, I was dealing
with risk premium models, for insurance companies, which were linear
combinations of an empirical part and an expected value related to some
prior guess---they identify the so-called credibility premium. I
thought to revisit the problem coherently with the predictive
distributions generated by an exchangeable sequence and asked myself
what the underlying de Finetti measure was: it turned out to be the law
of a Dirichlet process. I wrote this paper while I was working with
Cifarelli on the distribution of linear functionals of the Dirichlet
process. I then presented it at a conference, where Luciano Daboni, an
editorial board member of \textit{GIIA}, was present: he liked the paper a lot,
invited me to give a seminar in Trieste and proposed for me to publish
it in \textit{GIIA}. Some years later the same result was independently obtained
by Albert Y. Lo (Lo, \citeyear{lo91}).

\textbf{I:} During the 1980s you were probably one of the few
statisticians in Italy who published their papers in international
journals. Do you have any idea why this happened at the time?

\textbf{E:} Well, first of all, most people, both in statistics and
probability, did not even try to submit their work abroad. It was
simply not necessary for the progress in academic careers. Even many
mathematicians only published in Italian journals. Overall, the need
for trying to spread one's own work at an international level was not
felt yet. Actually, it was probably not even felt in the
Anglo-American world: it just happened that their journals then became
the ``international'' ones. By the way, papers that were published in
Italian journals with a very limited spreading were not all necessarily
of bad quality. On the contrary, some of them are very well known even
abroad and contain innovative ideas. Anyhow, in recent years things
have changed substantially and young researchers submit their work to
the best international journals.

\textbf{A:} We, as students, have nice memories of summer schools
organized by the Italian scientific community to support the spreading
of probability and statistics. You have been an active part of this
initiatives.

\textbf{E:} After attending some of them as a student, I~have been
involved several times in organizing and teaching at summer schools
that took place in various beautiful locations in Italy, such as
Cortona, Perugia, Livigno and Rh\^{e}me-Notre Dame. In addition to
being an opportunity to meet talented students, summer schools also
allowed me to get in contact, and actually build up friendships, with
some authoritative scholars such as Alan Agresti, Patrick Billingsley,
Albert Y.~Lo, Slava Sazonov, Henry Teicher and Jon Wellner.
Unfortunately, the cuts operated through the years by the Italian
governments have made it more difficult to sustain the organization of
such praiseworthy initiatives.

\textbf{A:} You then also started to collaborate with Sazo\-nov. In fact,
one of your articles we really enjoyed reading, for the wealth of
results and techniques it offers, is Regazzini and Sazonov
(\citeyear{regsaz00}). How did you convince him to do research on
Bayesian statistics?

\begin{figure}
\includegraphics{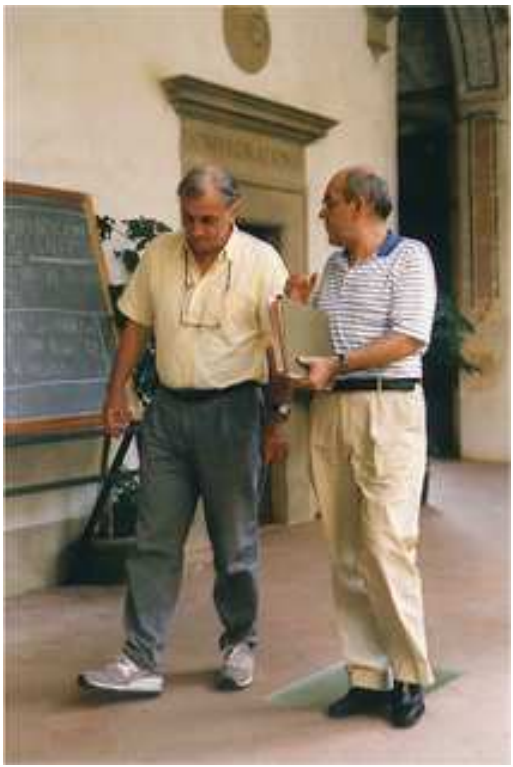}
\caption{Patrick Billingsley with Eugenio in Cortona, Summer School, 1989.}\label{f1}
\end{figure}

\textbf{E:} Slava was a wonderful person I really miss. He was a loyal
friend to me and an extraordinary mathematician. I first met him at a
conference at the beginning of the 1990s and, then invited him to
deliver a course, jointly with Albert Lo, at a summer school organized
by Bocconi University in 1992. He then taught also in the 1993 and 1994
editions. We started collaborating in 1996, while he was teaching a
course on ``Probability Theory in Hilbert spaces'' at the Italian
National Research Council in Milano. Our first joint work concerned
central limit theorems for partially exchangeable arrays of random
elements taking values in a Hilbert space. At the time I was also
preparing my lectures for a Ph.D. course on Bayesian nonparametrics to
be taught in Roma and I was dealing with the problem of estimating
a~statistical model by means of a mixture of Dirichlet processes. Such
a~problem was suggested by Diaconis and Ylvisaker (\citeyear{dy85}): with
Slava we showed that it is possible to construct a mixture of laws of
Dirichlet processes that approximates the distribution of any random
probability measure, with respect to the topology of weak convergence.
And we have been able to obtain, under suitable assumptions, the
corresponding approximation bounds for the posterior measures. These
results were presented at the 1st Workshop on Bayesian nonparametrics
that took place in Belgirate (Italy) in 1997. While we were working on
this paper, my mother became seriously ill and Slava has been very
important in supporting me in such a difficult period.

\begin{figure*}
\includegraphics{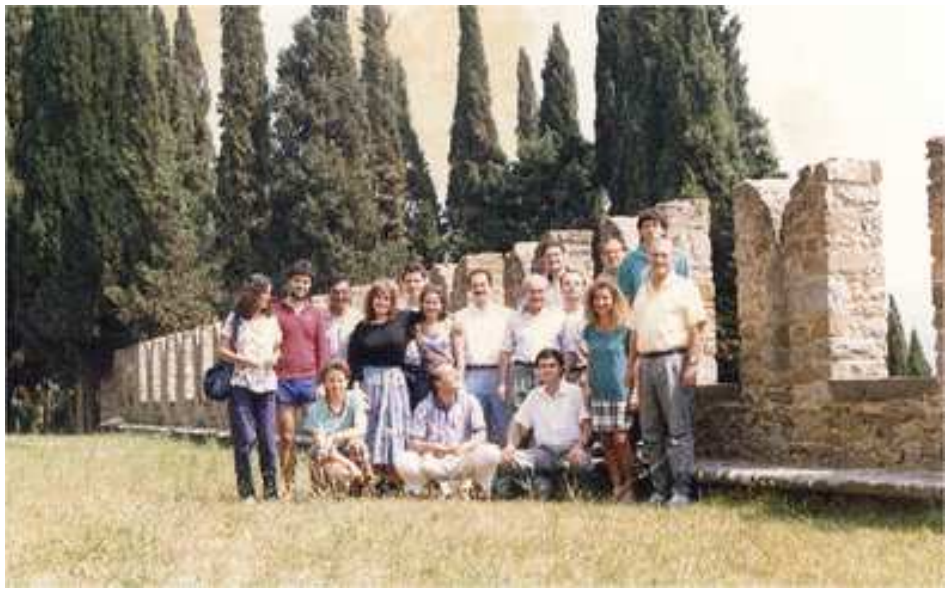}
\caption{Patrick Billingsley, standing in the far right, and Eugenio with some students in Cortona, Summer School, 1989.}\label{f2}
\end{figure*}

\textbf{I:} You have always had good relationships with probabilists
and statisticians from Russia. For example, Ildar Ibragimov is another
good friend of yours who has been several times in Pavia contributing
to the Ph.D. program. I had the pleasure to attend his lectures and
really enjoyed them.

\textbf{E:} It was actually Slava who suggested I contact Ildar
Ibragimov. In fact, I had asked Slava indications for possible
instructors for Ph.D. courses. And Slava mentioned about Ildar and told
me that in addition to being a great scientist he was an excellent
teacher. Of course, I knew Ildar by fame and I feared he would have not
accepted my invitation but he did. I got the chance to meet in person
not only a brilliant mathematician but also a wonderful person. His
courses in Pavia were greatly appreciated and I liked the fact that he,
and also Slava, was trying to adapt his lectures to the students'
background. Our Ph.D. classes are quite composite, with most students
having either mathematics or economics degrees. The former typically
have good backgrounds in pure maths but not in statistics and
probability, whereas for the latter it is the opposite. I remember
Ildar asking me, ``How come I have students in my class who know about
Radon--Nikod\'{y}m derivatives of stochastic processes but struggle with
Fourier transform coefficients?''

\section{Research}

\subsection{Bayesian Nonparametrics}

\textbf{A:} Your papers with Cifarelli on functionals of the Dirichlet
process (Cifarelli and Regazzini, \citeyear{cifreg79}, \citeyear{cifreg90}) are probably your most well-known
contributions to Bayesian nonparametrics. And it is amazing how many
connections your results have with a variety of research areas such as
combinatorics, mathematical physics, theory of stochastic processes,
the moments problem, and so on. Were you aware of these?

\textbf{E:} As I said, our original problem was merely of a~statistical
nature. From an analytical point of view, the task we were facing was
very challenging, but we were not aware of the connections with
seemingly unrelated areas of mathematics. We learned about some of
these relations thanks to the paper by Persi Diaconis and Johannes
Kemperman that was presented at the Valencia meeting in 1994; see
Diaconis and Kemperman (\citeyear{dk96}). In addition to embedding the
whole problem in a wider mathematical context, it is also very well
written and sketches a~few open problems; I strongly recommend reading
it. It is also thanks to this very same paper that my work with Cifarelli
gained some popularity.

\textbf{I:} The basic trick you resorted to was the inversion of a
Cauchy--Stieltjes transform for the mean of the Dirichlet process. How
did you arrive to this intuition?

\textbf{E:} The procedure actually relied on the determination of
recursive relations for the moments of the linear functional. Such a
strategy was inspired by the work of M.~Kac who used it to obtain the
well-known Feynman--Kac formula; see, for example, Kac
(\citeyear{kac49}). This closeness is further revealed by the adoption,
in our paper, Cifarelli and Regazzini (\citeyear{cifreg79}), of the
same symbols used by Kac! Cifarelli had successfully used it to
establish a closed form expression for the probability distribution of
the integral of the absolute value of the Brownian bridge in Cifarelli
(\citeyear{cif75}). These recursive relations we obtained allowed us to
determine the Laplace transform whose iteration yields the
Cauchy--Stieltjes\break transform.\vadjust{\goodbreak} We then resorted to the inversion formulae
of the Cauchy--Stieltjes transform to deduce an exact form for the
probability distribution of a linear functional of the Dirichlet
process. Most of these ideas were already contained in Cifarelli and
Regazzini (\citeyear{cifreg79}). In Cifarelli and Regazzini
(\citeyear{cifreg90}) we basically completed that paper and provided
some further insight.

\textbf{A:} More recently you developed an alternative method based on
an inversion formula for the characteristic function.

\textbf{E:} The approach you are referring to was inspired
by the representation of the Dirichlet process as the normalization
of a gamma process that was first pointed out by Ferguson himself in
his 1973 paper.
This representation combined with a suitable inversion formula led to new
forms for the probability distribution of the mean of a Dirichlet process,
which are recorded in a paper with Alessandra Guglielmi and Giulia Di Nunno.
I~have then extended, with the two of you, the approach to deal with
means of random
probability measures induced by the normalization of a generic process
with independent increments.
%See \cite{rlp}.

\begin{figure*}
\includegraphics{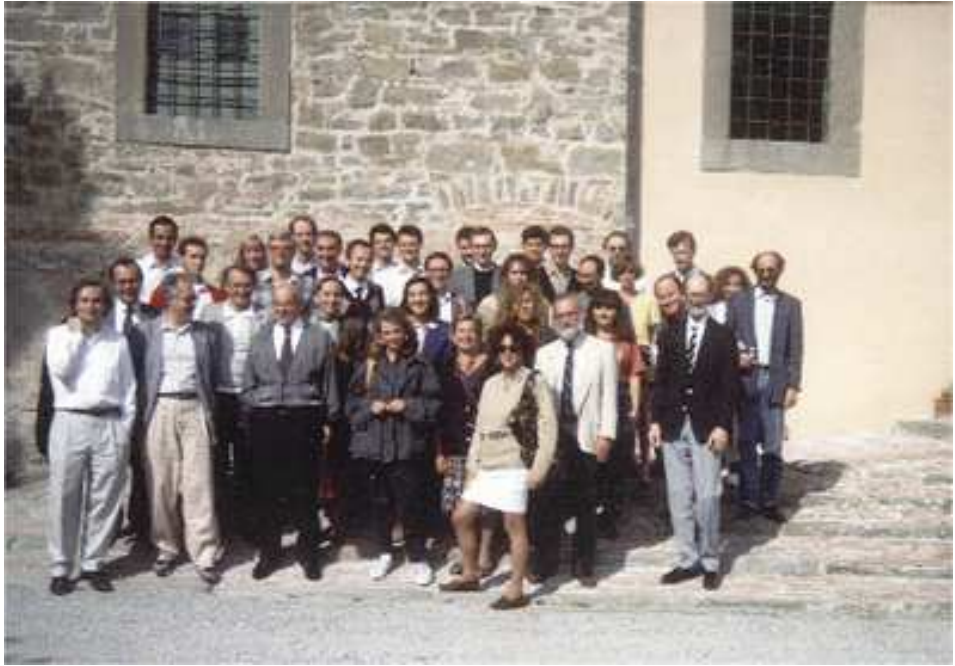}
\caption{Workshop on ``Recent developments in exchangeability,'' Cortona, October 1991. Among others,
  Luigi Accardi, Donato M. Cifarelli, Guido Consonni, Persi Diaconis, Joe Eaton, Colin Mallows,  Jan von Plato,
  Maurizio Pratelli, Wolfgang Runggaldier, Marco Scarsini, Brian Skyrms, Fabio Spizzichino, Piero Veronese,
  Wolfgang Woess and Eugenio.}\label{f3}
\end{figure*}

\begin{figure}
\includegraphics{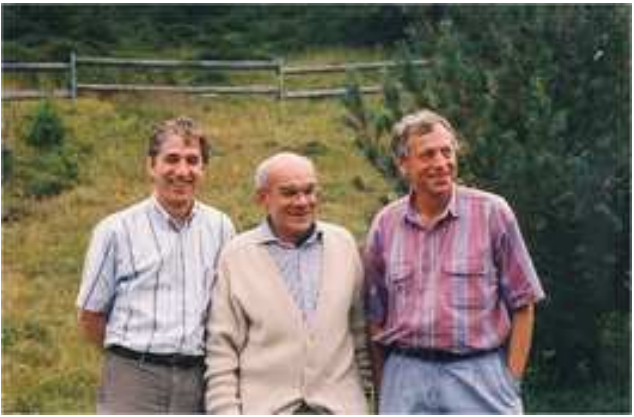}
 \caption{From the left: Alan Agresti, Eugenio and Slava Sazonov in Livigno, Summer School, July 1993.}\label{f4}
 \vspace*{-3pt}
\end{figure}

\textbf{I:} At the moment, Bayesian nonparametric regression is a hot
topic. In this respect, a paper of Cifarelli and yourself has been
recently ``rediscovered'' (Cifarelli and Regazzini,
\citeyear{cifreg78}). Can you talk to us about its origin and contents?

\textbf{E:} The original goal of our research was to determine a
probability distribution for partially exchangeable arrays of random
elements. In particular, we were looking for a solution that could be
treated analytically, while avoiding the independence assumption among
rows. These were the two reasons which led us to the idea of resorting
to the mixture of products of Dirichlet processes. We have been able to
determine the associated system of predictive laws and the distribution
of vectors of functionals. In a~parametric setting, partial
exchangeability had been incorporated in a paper by Lindley and Smith
(Lindley and Smith, \citeyear{smith}). I then used our model to~study
credibility\vadjust{\goodbreak} formulae with collateral data. Cifarelli had also developed
the model for applications to ANOVA and linear models, the latter in
collabora\-tion with Marco Scarsini and Pietro Muliere. We~did not even
submit the paper to a journal, since, as I~said, at the time a
technical report or a journal publication counted the same for us.
Nowadays, I~am really pleased to see the recent explosion of propo\-sals
on dependent nonparametric models, somehow in the spirit of our 1978
paper, developed by S.~Mac\-Eachern, P.~M\"{u}ller, D.~Dunson and many
others.\looseness=-1

\begin{figure*}
\includegraphics{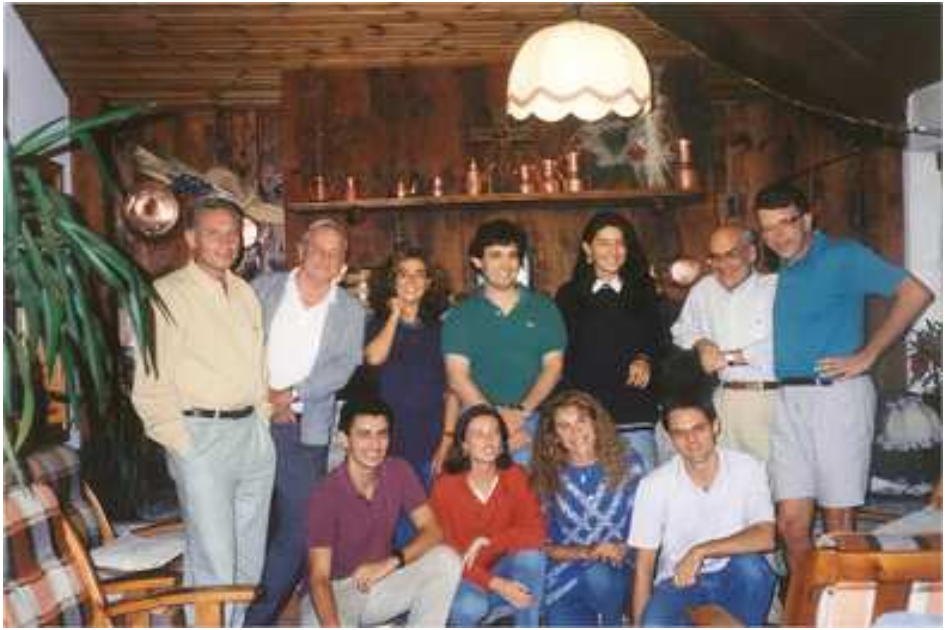}%
\vspace*{-2pt}
\caption{Slava Sazonov and Patrick Billingsley (on the left) and Eugenio and Andrew Rukhin (on the right) with some students in Rhemes-Notre-Dame, Summer School, July 1994.}\label{f5}
\vspace*{-4pt}
\end{figure*}

\textbf{I:} In our opinion, the work of two probabilists, John F. C.
Kingman and Jim Pitman, has to be listed among the main and most far
reaching contributions to Bayesian nonparametrics, even if not directly
focused on it. Do you share this view?

\textbf{E:} I am strongly in favor of a Bayesian approach that solely
relies on the specification of distributions for observable random
elements. Therefore, in general, I like all those contributions and
tools that aim at providing systems of predictive distributions related
to modeling and applications. These do not resort to conditional
distributions, given parameters (either finite or
infinite-dimensional) that in some applications would be devoid of any
empirical meaning. And, the works by Kingman and Pitman, although
originated in different research areas, have an\vadjust{\goodbreak} important impact on
Bayesian statistics. Even though I read their papers only recently, I
have appreciated them very much since they open up the possibility of
implementing the Bayesian paradigm in the direction I lean toward.

\begin{figure}[b]
\vspace*{-3pt}
\includegraphics{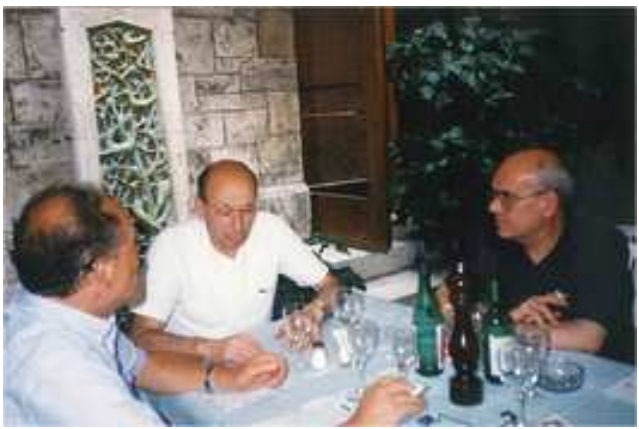}%
\vspace*{-2pt}
\caption{From the left: Giorgio Dall'Aglio, Henry Teicher and Eugenio in Perugia, Summer School, August 1995.}\label{f6}
\end{figure}

\subsection{Exchangeability}

\textbf{A:} The contributions of Kingman and Pitman you just mentioned
are closely related to exchangeability, a~topic you extensively worked
on both from a~statistical and probabilistic point of view.

\begin{figure}[b]
\vspace*{-4pt}
\includegraphics{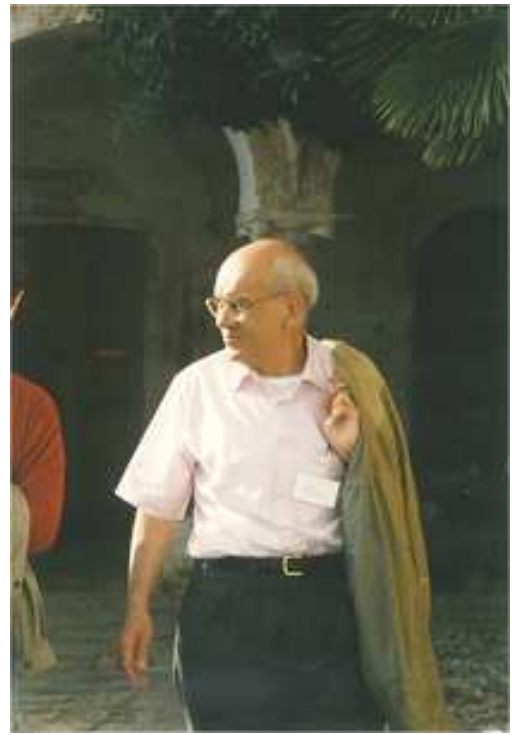}%
\vspace*{-2pt}
\caption{Eugenio in Belgirate, 1st Bayesian Nonparametrics Workshop, June 1997.}\label{f7}
\end{figure}

\begin{figure}
\includegraphics{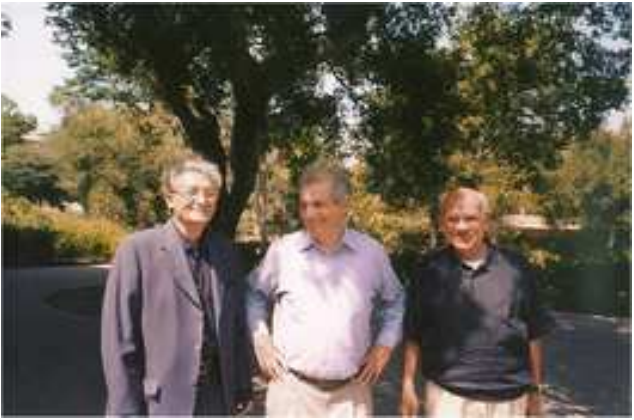}%
\vspace*{-2pt}
\caption{From the left: Donato M. Cifarelli, Persi Diaconis and Eugenio at Stanford University, July 2002.}\label{f8}
\vspace*{-4pt}
\end{figure}

\begin{figure*}[b]
\vspace*{-4pt}
\includegraphics{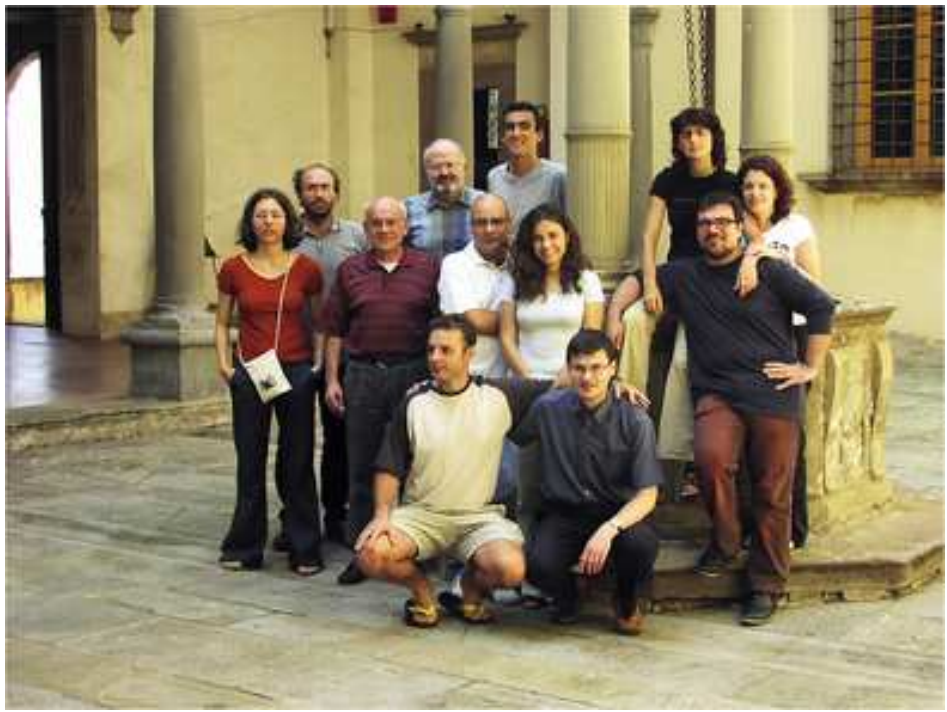}%
\vspace*{-2pt}
\caption{Jon Wellner and Eugenio with some students in Cortona, Summer School, August 2004.}\label{f9}
\end{figure*}

\textbf{E:} My interest in exchangeability was stimulated by reading
de~Finetti's papers. The first place where I came\vadjust{\goodbreak} across the statement
of de~Finetti's representation theorem was the monograph by Lo\`{e}ve.
But I~could not understand its statistical implications. I~could
appreciate its relevance for inductive reasoning only\vadjust{\goodbreak} through a careful
study of de~Finetti (\citeyear{defin30}, \citeyear{defin37}): in my
opinion, these papers really stand out in terms of conceptual and
mathematical rigor and effectiveness in highlighting the role of
exchangeability for induction, and remain unbeaten to date. Of course,
the modern uses of exchangeability and the key role it plays in
modeling a variety of phenomena are probably beyond what de~Finetti
could have expected.

\textbf{I:} You have also been working on characterization theorems in
this context.

\textbf{E:} You are probably referring to results I have obtained with
Sandra Fortini and Lucia\vadjust{\goodbreak} Ladelli and that characterize systems of
predictive distributions associated with exchangeable sequences of
random elements. I have also noted that these kinds of results have
recently attracted more and more interest in Bayesian nonparametrics
practice. Another interesting characterization was obtained in a paper
I coauthored with Giovanni Petris where we dealt with exchangeability
in the presence of finitely additive probabilities: we stated and
proved a weak version of the representation theorem that reduces to the
celebrated de~Finetti theorem (strong version) if one specializes to
the case of $\sigma$-additive probabilities. In this situation, we were
also able to use the representation theorem to show existence of
a~random probability measure defined by means of a~system of
finite-dimensional distributions agreeing with Ferguson's framework.

\textbf{I:} You have also provided nice contributions to the
investigation of properties of partially exchangeability.

\begin{figure*}
\includegraphics{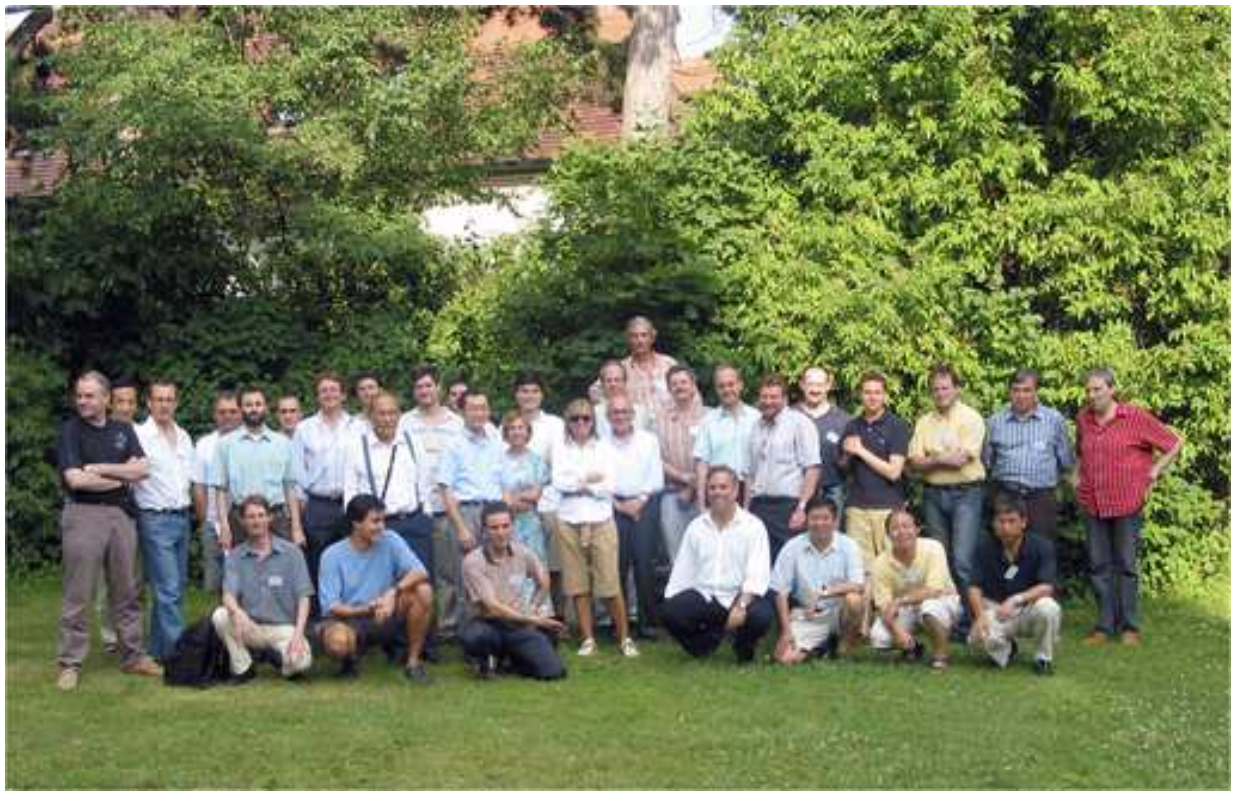}
\caption{Eugenio at a conference on ``Non--linear PDEs: homogeneization and kinetic equations,'' Wien, June 2006. In the picture, among others: Peter Markowich, Pierre Degoud, Eric Carlen, Maria C. Carvalho, Ester Gabetta, Giuseppe Toscani, Cristian Ringhofer, Anton Arnold and George Zubelli.}\label{f10}
\end{figure*}

\textbf{E:} Indeed, I have been, and I still am, interested in forms of
dependence more general than exchangeability, as witnessed by some
contributions I have already mentioned before, such as the paper on
mixtures of products of Dirichlet processes or the formulation of a
central limit theorem for partially exchangeable arrays. Besides these,
I wish to mention a nice characterization\vadjust{\goodbreak} of partially exchangeable
arrays that has been established in a paper I~wrote with Fortini,
Ladelli and Petris. Indeed, we proved a conjecture formulated in
de~Finetti (\citeyear{defin59}), according to which a suitable random
matrix related to the transitions of a recurrent process is partially
exchangeable if and only if the law of the process can be represented
as a mixture of laws of Markov chains. Moreover, we have been able to
show that de~Finetti's definition of partial exchangeability is
equivalent to the one provided by Diaconis and Freedman in a couple of
papers they wrote in 1980.

\subsection{Subjective Probability}

\textbf{I:} In some of your work %(\cite{reg87}, \cite{brr91,brr98})
you have also provided some insight into an approach to Bayesian
statistical inference based on finitely additive conditional
probabilities.

\textbf{E:} I started getting involved into research on fini\-tely
additive conditional probabilities after reading some papers by
R.~Scozzafava in the first half of the 1980s. In fact, I grew convinced
that countable additivity was not justifiable---as a necessary
condition---unlike finite\vadjust{\goodbreak} additivity which is necessary for the
validity of de~Finetti's coherence principle.
%even from a frequentist point of view, a fact Kolmogorov was well
%aware of (see \citet[Ch.~2, Sec.~1]{kolmo33}).
Therefore, finitely additive probabilities have to be considered as
admissible and I became interested in revisiting known results in
probability as particular cases of the finitely additive framework. In
particular, I found the interpretation of the definition of conditional
probability, as given by Kolmogorov, unsatisfactory. Conditioning is
based on classes of events that partition the whole sample space and
that become finer and finer: conditional probability is then obtained
through a limiting process in terms of a Radon--Nikod\'{y}m derivative,
and depends on the class of events one conditions on. In de~Finetti's
approach, a conditional probability, given an event, is defined through
a natural, and unavoidable, strengthening of the coherence principle.
De~Finetti himself had hinted at such a possibility, without developing
his idea in general mathematical terms. I tried to make this more
explicit in some papers I wrote during the 1980s in Regazzini
(\citeyear{reg85}, \citeyear{reg87}). These topics have been object of further investigation
by my friends P. Berti and P. Rigo. An important\vadjust{\goodbreak} point is that many
situations that appear as paradoxical if one refers to Kolmogorov's
conditional probabilities can be justified within the finitely additive
framework.

\begin{figure}
\includegraphics{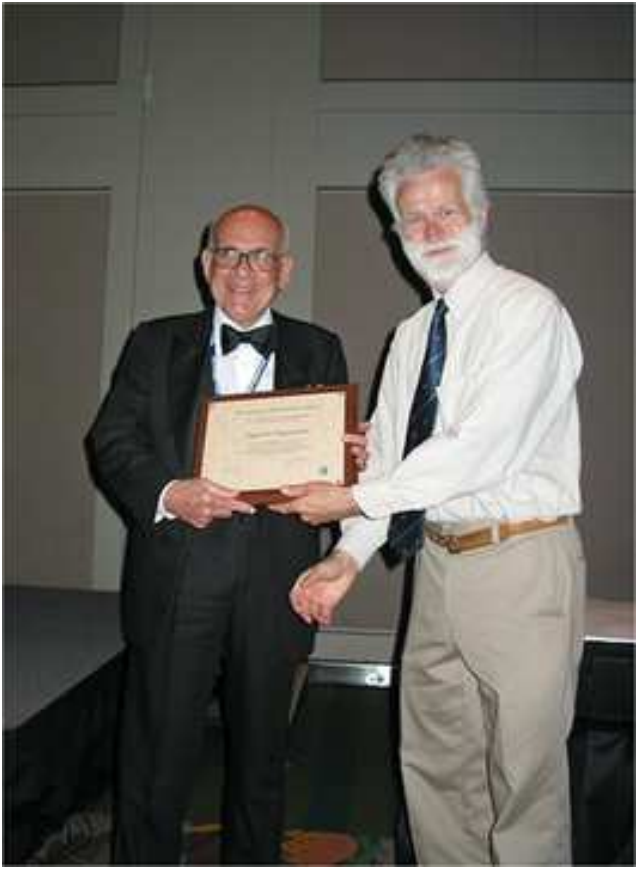}
\caption{IMS President Jim Pitman with Eugenio at the IMS Fellows Ceremony, 70th IMS Annual Meeting, Salt Lake City, July 2007.}\label{f11}
\vspace*{-3pt}
\end{figure}

\textbf{A:} Can you provide us with an example?

\textbf{E:} The most well-known is probably Borel's paradox. Indeed, if
a uniform distribution on the surface of a sphere is defined, with
respect to a specific choice of geographic coordinates (namely,
latitude and longitude), one would expect that the conditional
distribution for latitude, given a fixed longitude, is uniform.
However, this does not happen in Kolmogorov's framework. In
de~Finetti's approach, instead, one can adopt the more intuitive
probability assessment even if it would be nondisintegrable.
%if one wishes to preserve both coherence and conglomerability
%properties, conditional uniformity does not hold true even if
%intuitively it should.
The reason for such a behavior can be traced back to the specific
notion of conditional probability according to Kolmogorov's approach,
since it does not admit the evaluation of the probability of an
``isolated event'' with probability zero. On the contrary, de~Finetti's
setup is open to different solutions: indeed, disintegrability turns
out to be \textit{not necessary} for coherence.\vadjust{\goodbreak}

\textbf{I:} Another amusing aspect of finitely additive conditional
probabilities emerges from your work on well-calibration of systems of
predictive distributions.

\textbf{E:} Loosely speaking, well-calibration corresponds to
situations where the distance between weighted averages of forecast
probabilites and empirical observations converges to zero as the number
of observations, and forecasts, increases. Kolmogorov's theory always
yields well-calibrated predictions or forecasts. This corresponds to a
somehow unrealistic situation in practice, since one would also expect
cases of not well-calibration. With P.~Berti and P.~Rigo we were
interested in checking whether the same was true within de~Finetti's
theory as well. Our curiosity to this problem was stimulated by a paper
of Phil Dawid (Dawid, \citeyear{daw82}). The answer we got was
naturally affirmative for \textit{strategic} conditional probabilities.
The term strategic was coined by Dubins and Savage in their well-known
monograph where they resorted to de~Finetti's theory to solve quite
complicated measurability problems. Strategic conditional probabilities
do indeed preserve, in a finitely additive setting, the
disintegrability property that characterizes Kolmogorov's definition.
As for well-calibration, we were able to show that, beyond strategic
evaluations, there exist not well-calibrated coherent Bayesian
predictors with positive probability.\looseness=1

\textbf{A:} Many critics of de~Finetti's subjectivistic standpoint in
probability theory use, as an argument for supporting their position,
de~Finetti's sentence ``probability does not exist.'' What can be
replied to such objections?

\textbf{E:} First of all, one should consider the provocative nature of
de~Finetti's sentence. Moreover, its meaning should not be
decontextualized. According to de~Finetti, if one wants to give
probability an objective meaning, one should prove its existence. In
other words, there should be an existence theorem, a clear proof of the
existence of an object termed ``probability.'' For example, the
interpretation of probability as a limiting frequency cannot be
considered as a proof, even if just empirical, of its existence. Hence,
he used the expression ``probability does not exist'' just to make the
point that probability has simply a subjective meaning. It is also to
be said that most of the criticism raised against subjectivism
basically refers to the contents of his two-volume monograph,
de~Finetti (\citeyear{defin70}), which is, according to its subtitle,
``a critical introductory treatment.'' In my opinion, de~Finetti's
position can be better discussed by relying on his early works,\vadjust{\goodbreak} which
are more concise, go straight to the point and display more
mathematical and formal details.

\textbf{I:} A noteworthy scholar who contributed to the theory of
finitely additive probabilities was Lester Dubins. You were also a good
friend of his and had the chance to host him in Milano.

\textbf{E:} Dubins had been in Italy several times and he delivered
courses at summer schools. He was very fond of Italy and, in the second
half of the 1980s, I invited him once to stay for a month in Milano. We
had discussions on various research topics. He was the source of many
ideas that I later developed in my research. In those years I was
mainly working on technical aspects of nonparametric inference, whereas
he could provide me with many insights into theoretical issues related
to finite additivity that turned out to be of great importance to me.

\subsection{Probabilistic Methods for Mathematical Physics}

\textbf{A and I:} You have lately become interested in some problems in
mathematical physics. How did it happen?

\textbf{E:} Moving to Pavia in 1998, I joined a Mathematics Department
with a few internationally well-known mathematical physics scholars. I
started interacting with them and at some point a colleague of mine,
Ester Gabetta, showed me some papers where the Central Limit Theorem
was used to describe the convergence to equilibrium of the solution of
certain kinetic equations. In particular, I read two papers, McKean
(\citeyear{mckean66,mckean67}), that spurred my enthusiasm for the
topic. I tried to understand and extend the connections with
probability, and could count on the collaboration of my colleagues to
help me understand the problem from the perspective of physics.
Furthermore, the encouragement from Eric Carlen and Maria Carvalho has
been important for pursuing my research in this direction. In fact, they liked our first results
and suggested us to publish them (see Gabetta and Regazzini, \citeyear{gr2006}): there we
obtained some identities that came in handy for later developments of the work in
this area.

\textbf{A and I:} Was this line of research as rewarding as others you
have pursued in your career?

\textbf{E:}
I would say I am happy about what I have achieved so far with my coauthors.
Starting from the Kac model, which is generally considered as a~toy
model, we obtained some interesting results concerning the characterization
of the initial data in order to gain\vadjust{\goodbreak} convergence to equilibrium. We
have also
considered situations where the energy, interpreted as the variance of
the initial
datum, is infinite and we performed an analysis of the speed of
convergence. In
these studies, I have also collaborated with Lucia Ladelli and Federico
Bassetti. Later,
%while developing some published results on the speed of convergence,
%in total variation, of the solution of the Kac equation,
I have supervised the thesis of Emanuele Dolera, a Ph.D. student in
Pavia. This work has required a strong effort that was rewarded by the
achievement of a noteworthy result proving the validity of a conjecture
formulated in the 1966 McKean paper. In the last 40 years many scholars
have worked hard with the aim of proving it.

\section{Thoughts on Foundational Issues and Research in Statistics}

\textbf{I:} In some of the previous questions we have lingered on the
subjectivistic interpretation of probability. What is the most relevant
impact this has on statistics?

\textbf{E:} A crucial point to understand is whether it is worth
preserving an axiomatization based on countable additivity. Of course,
I think it does not generally have a statistical justification that
makes its use necessary. If finitely additive probabilities are also
admissible, then a considerable number of results in the literature
should be revisited. I have already mentioned that one should
reconsider the definition of conditional expectation. Moreover, a
number of limiting theorems should be reformulated in order to account
for this more general framework. These issues are also of great
relevance in statistics regardless of the approach, either frequentist
or Bayesian, one adopts.

\textbf{A:} Does this lead, among others, to a rethinking of Bayesian
procedures?

\textbf{E:} Indeed, Bayesian procedures are typically implemented by
assuming complete additivity and this leads to assume some of its
implications as necessary. Let us consider, as an example, the
Dirichlet process. A~well-known result is that the Dirichlet process
selects, almost surely, discrete probability measures. However, such a
property holds true for the countably additive extension of the
collections of finite-dimensional probability distributions of the
process. There are other non $\sigma$-additive extensions for which the
Dirichlet process selects nondiscrete distributions with positive
probability. This points to the fact that in statistical practice one
should avoid assessing a probability for objects devoid of empirical
evidence. For example, take the proposition stating that de~Finetti's
measure is the law of the (almost sure) weak limit of the empirical
distribution: thus, it depends on infinitely many observations and
concerns ``transcendent''---in de~Finetti's words---conditions not
directly verifiable. The conclusion of such a proposition could be
obviously false with non $\sigma$-additive extensions. On the other
hand, the fact that de~Finetti's measure is the weak limit of the low
of the
empirical distribution, as the sample size increases, is, in any case,
true: in my opinion this suffices with respect to sound statistical
goals. I think this is an important foundational aspect, which is often
neglected and should be further investigated.

\textbf{I:} Are you saying that one should have clear in mind the
different levels at which mathematics and statistical applications
operate?

\textbf{E:} More or less, that is what I mean. Indeed, it is true that
mathematics
makes parameters interpretable as limits of (or of functionals of)
empirical processes, but it does not automatically grant that inference
on them are legitimate.
%On the other hand, the representation theorem is still useful, from a
%statistical standpoint, since it involves weak convergence and the
%limit law can be legitimately used as an approximation.

\textbf{A:} Does this position contrast with the usual way of
presenting a Bayesian model as the combination of a likelihood and a
prior?

\textbf{E:} Let me start by making an important point that reflects my
view on statistics: if inference is seen as a decision problem to be
solved under uncertainty and if one agrees that probability is a tool
to resort to, then there is no other choice but the Bayesian approach.
Nonetheless, I agree with what Diaconis and Ylvisaker say at the
beginning of their paper Diaconis and Ylvisaker (\citeyear{dy85}):
Bayesian statistics cannot be reduced to the elicitation of a prior and
the automatic application of Bayes' theorem. Hence, I would give an
affirmative answer to your question if one conditions on unobservable
quantities. But this is not limiting the scope of Bayesian inference at
all. Indeed, one can think of inferential procedures that can still be
implemented in this more general framework, even when unobservable
parameters are involved. The previously mentioned ``weak''
interpretation of the de~Finetti measure says that a prior distribution
can be viewed, in any case and with no distinction between observable
and not observable parameters, as an approximation of the law of
a~frequency distribution or of some functional of it. Moreover,
prediction can be carried out without relying on the Bayes--Laplace
paradigm: it is enough to specify the system of predictive
distributions connected to the exchangeable sequence. And I have
appreciated very much the work by J.~Pitman which, in the spirit of
de~Finetti's stance, relies on the proposal of systems of predictive
distributions that are then proved to be associated to an exchangeable
sequence.

\textbf{A:} I also guess that a subjectivist would not agree on the
notion of posterior consistency as a frequentist validation criterion
of Bayesian nonparametric methods.

\textbf{E:} %I have to admit that my opinion on this is strongly
%influenced by my subjectivistic view of probability.
I have to admit that, besides the Bayesian context, I am skeptical on
the use of consistency in a~frequentist setting as well. On the one
hand, these limiting results are very neat and beautiful from
a~mathematical point of view. But, on the other, they lack a sensible
statistical interpretation. This is very well discussed in de~Finetti
(\citeyear{defin70}), Volume~2, in the section devoted to the laws of
large numbers where he motivates why results, such as consistency, do
not represent justifications of statistical procedures under the
assumption of stochastic independence. The same can be said for the
Glivenko--Cantelli theorem. Of course, my position encompasses commonly
used frequentist validation criteria adopted in a Bayesian framework. A
different role must be attributed to approximation results, like
Central Limit Theorems or, also, the ``weak'' interpretation of the
de~Finetti measure, for which these concerns do not apply.

\textbf{I:} So, what are the kind of asymptotic problems that you think
are interesting for Bayesians?

\textbf{E:} The kind of results I like are those in the spirit of
Blackwell and Dubins (\citeyear{blackdub62}) where they investigate the
phenomenon of the merging of opinions. This is a very nice finding both
from a mathematical and from a statistical point of view. On the one
hand, it is a general result valid even beyond exchangeability. On the
other hand, it has a nice statistical interpretation for Bayesians
since it hints at the predominance of empirical findings over different
subjective prior opinions as the sample size or, in other terms, the
amount of information, increases. Other results of great interest are
those that currently are designated as Bernstein--von Mises type
theorems for the posterior distribution. It is worth noticing that
among the first contributions to this topic there is also an important
paper, Romanovsky (\citeyear{roman31}), published on \textit{GIIA}.

\textbf{A:} From what you said up to now on the interplay between
Bayesian inference and de~Finetti's interpretation of probability, a
valuable research topic would focus on the analysis of the asymptotic
behavior of the predictive distributions.

\textbf{E:} You are right. In my opinion, an important issue to address
is the analysis of the distance between the predictive and the
empirical distributions. Instead of looking at the limiting behavior,
it would be more interesting to analyze how such a discrepancy changes
for any sample size $n$ and, \textit{a fortiori}, as $n$ increases.
Since the predictive can also be obtained as a functional of the
posterior distribution, one can also gain some insight if one relies on
convergence theorems, which say that the posterior converges, in some
sense, to a distribution concentrated on the limit of the empirical
process. In this respect, Bernstein--von~Mises type results are useful.

\textbf{I:} You have had a large number of students, and by now also
descendants, working in many different universities in Italy and
abroad. In your opinion, what is the background a statistics student
needs to perform well in nowadays research and what are the topics you
would suggest to pursue?

\textbf{E:} As I said earlier, I see statistics as inductive reasoning
under the supervision of probability theory. Therefore, it is natural
that I firmly believe that statisticians should have a solid background
in probability: the more the better. However, a statistician must also
be able to think through the logical and philosophical aspects of what
she/he is doing. This concerns modeling, the understanding of practical
implications yielded by the mathematical formulation that is used, and
the interpretation of results. Mathematical skills are not enough,
logical and conceptual rigor being a necessary complement. One needs to
be able to handle statistics since it is a~powerful instrument, which
allows one to make substantial steps forward, compared to traditional
deterministic procedures. Statistics can get you close to the best
solutions, avoiding overwhelming technical and mathematical
difficulties that often arise within deductive deterministic reasoning.
The latter approach lacks the flexibility of a learning mechanism,
whereas in the probabilistic framework everything is kept under
control: you have a law which governs everything and, unless you change
the learning mechanism, it allows one to learn from experience in a way
that is transparent and controlled by Bayes' theorem.

\textbf{A:} In modern science the specialization of re-\break searchers is
constantly increasing. Even probability and statistics, which have
grown in close relationship to each other, seem to be drifting apart.

\textbf{E:} You can observe the fragmentation of fields all over the
place. This phenomenon also originates from an excessive\vadjust{\goodbreak} specialization
that characterizes most undergraduate studies. The situation was, in
the past, quite different and there were many scholars with a wide
spectrum knowledge and diversified cultural and scientific interests.
De~Finetti and Gini are excellent examples in this respect. That said,
fragmentation in research is unavoidable and it would be unrealistic to
try reversing it. It is just a pity to see that it tends to create
duplications and repetitions, whereas a more cohesive scientific
community could produce better results in a collective effort. In
statistics, Bayesian statisticians have kept to themselves for some
time in reaction to the then mainstream statistics, which was certainly
not in favor of Bayesian methods. Now with Bayes statistics
well-established, I note that younger generations are more open to
interactions with non-Bayesian, which in my opinion is certainly
beneficial. A different issue is the specialization in education, which
should be contrasted to some extent because it precludes possible paths
to future researchers. As I have already said, every statistician
should have a solid background in probability and every probabilist
should know the basics of statistics, which is a noble and fascinating,
at least to me, field of application of probability.

\textbf{I:} How should, in your opinion, a good statistics paper be
structured?

\textbf{E:} Well, first of all the definition ``good'' is to be
considered with reference to the historical period. Until some years
ago theoretical papers were very appreciated, whereas nowadays applied
work plays an increasing role thanks to the computational tools.
However, I think that, in general, different forms of motivation are
equally valid: an enrichment of the available tools, an improvement
over other existing contributions or a useful application are all fine.
However, in all cases it is crucial that the paper is logically sound
and coherent with its motivation. This is essential since we write for
the scientific community and not for the general public, which is
another job. For instance, I do not like methodological papers, to
which an illustration has been evidently added only for editorial
needs. A methodological contribution can stand on its own if its
motivation is sound. While I was young I experienced some of the last
manifestations of Gini's school, which as a~rule of thumb required
publications to include data, a~table and a plot. To me this does not
make sense. I~also do not like applied papers in which one sets forth
a~model, analyzes a couple of data sets and concludes that the model
works well. Any so-called empirical\vadjust{\goodbreak} validation does not show anything
and is not enough to assess the suitability of a model. Indeed, there
should also be a sensible methodological motivation in the sense that
one should explain which features of a certain model make it more
appropriate for the problem at hand.

\textbf{A and I:} Moving away from statistics and probability, we
already mentioned your passion for music. How did you get fond of music
and what else are you interested in when you do not do research?

\textbf{E:} Being born in Cremona, my passion for music is quite
natural: it is 20~km away from the places Giuseppe Verdi grew up in,
melodrama is popular and there is a great tradition. It is also the
hometown of Claudio Monteverdi and of Amilcare Pon\-chielli, two famous
composers. Last but not least, it is the town of lute makers, the most
renowned being Antonio Stradivari. Even the general public knows opera
very well. Then, starting from opera when I was young, my interest
extended to symphonic music. I have also been fond of visual arts since
I was a kid: I loved paintings, architecture and sculpture since I
related them to Italian history. I remember having a great teacher at
school who used to emphasize links between history, arts and
literature. A peculiar feature of Italy is that, if you are interested
in any historical aspect, you necessarily end up considering also
painting, sculpture and architecture since they are all intimately
connected. We obviously benefited from Christian culture that played a
fundamental role, after the fall of the Roman Empire, in preserving the
wonders inherited from classical Greek and Roman traditions and in
promoting arts in forms we can today admire while visiting churches,
historical buildings, squares and museums. Moreover, during the
Renaissance, there were a large number of small states and many of them
had patrons who liked and could afford being surrounded by artists.
Hence, many towns developed their peculiar artistic heritage.

\textbf{A and I:} Eugenio, thanks a lot for patiently answering all our
questions.

\textbf{E:} Thanks to you for listening to all this!

\section*{Acknowledgments}
Many thanks to Elena Di Biase for a careful reading of the manuscript
and several useful suggestions which improved the presentation. This
work is partially supported by MIUR, Grant 2008MK3AFZ.

% imsref loaded by arune.pranskunaite, 2011-06-21 13:23:31
%

\end{document}